\documentclass[aps,superscriptaddress,nofootinbib,eqsecnum,pra,notitlepage,twocolumn]{revtex4-2} 

\usepackage[caption=false]{subfig}
\usepackage{cancel} 
\usepackage{amsfonts}
\usepackage{amsthm,bm}
\usepackage{amsmath}
\usepackage{amssymb}
\usepackage{graphicx,xcolor}
\usepackage{float}
\usepackage{hyperref}
\usepackage{dcolumn}
\usepackage{soul}
\usepackage{ulem}
\usepackage{verbatim}
\DeclareMathAlphabet{\mathpzc}{OT1}{pzc}{m}{it}

\begin{document}

\title{Momentum Distribution of a mass-imbalanced three-body system across the Efimov-Unatomic transition}

\author{D. S. Rosa}

\affiliation{Instituto Tecnol\'{o}gico de Aeron\'{a}utica, DCTA,
  12228-900 S\~{a}o Jos\'{e} dos Campos, SP, Brazil}

  \author{R. M. Francisco}

\affiliation{Instituto Tecnol\'{o}gico de Aeron\'{a}utica, DCTA,
  12228-900 S\~{a}o Jos\'{e} dos Campos, SP, Brazil}

  \affiliation{Université Paris-Saclay, CNRS/IN2P3, IJCLab, 91405 Orsay, France}
  
  \author{T. Frederico}

\affiliation{Instituto Tecnol\'{o}gico de Aeron\'{a}utica, DCTA,
  12228-900 S\~{a}o Jos\'{e} dos Campos, SP, Brazil}

\author{G. Krein}

\affiliation{Instituto de F\'isica Te\'orica, Universidade Estadual Paulista,
Rua Dr. Bento Teobaldo Ferraz, 271-Bloco II, 01140-070 S\~ao Paulo, SP, Brazil}

\author{M. T. Yamashita}

\affiliation{Instituto de F\'isica Te\'orica, Universidade Estadual Paulista,
Rua Dr. Bento Teobaldo Ferraz, 271-Bloco II, 01140-070 S\~ao Paulo, SP, Brazil}

\begin{abstract}
We investigate the single-particle momentum distribution, the two-body contact and three-body tail coefficients of mass-imbalanced three-body systems at the critical dimension $D_c$, where the transition between discrete and continuous scale invariance takes place as the spatial dimension is tuned between three and two dimensions. We show that the asymptotic momentum distribution at $D_c$ is governed by a distinct logarithmic scaling structure, which differs fundamentally from both the log-periodic behavior of Efimov states and the power-law scaling of the unatomic regime. This structure contains an additional quadratic-logarithmic contribution, characterized by the asymptotic coefficient $C_3''$, which leads to a finite and well-defined description of the momentum tail at the transition. The coefficient $C_3''$ depends sensitively on the mass imbalance, changing sign across different mass configurations and vanishing for identical particles. For heavy-light-light configurations, the subleading three-body term vanishes at a characteristic momentum, leaving the leading two-body contribution at the order retained in the asymptotic expansion. The physical accessibility of this cancellation depends on the separation between the trimer size and the interaction range. We further analyze the narrow intermediate region connecting the Efimov and unatomic regimes, here identified as an intermediate scaling regime, whose extent and properties are strongly controlled by the mass ratio.
These results establish the critical dimension as a regime with emergent scaling properties and provide quantitative momentum-space signatures for probing the transition between discrete and continuous scale invariance in few-body quantum systems.
\end{abstract}

\maketitle

\section{Introduction}

Few-body quantum systems with short-range interactions exhibit remarkable universal properties that are largely independent of the microscopic details of the interaction potential. A striking manifestation of this universality is the Efimov effect~\cite{efimov}, predicted by Vitaly Efimov in 1970, in which an infinite series of three-body bound states emerges when the two-body interaction approaches the unitary limit. A decisive step toward the experimental exploration of this phenomenon was provided by the realization of Bose–Einstein condensates in dilute atomic gases~\cite{BEC} and by the development of magnetic Feshbach resonances~\cite{feshbach}, which enable precise tuning of the two-body scattering length and allow cold-atom systems to reach the universal regime where Efimov states can be observed~\cite{efimovexp1,efimovexp2,efimovexp3}. These advances established ultracold atomic gases as a versatile platform for investigating universal few-body phenomena. Over the past decades, Efimov physics has developed into a broad research field spanning a wide range of physical systems and length scales, from ultracold atomic gases~\cite{naidonreview,greenereview} and halo nuclei~\cite{hammerreview} to molecular clusters~\cite{molecular1,molecular2} and other few-body quantum platforms~\cite{son2dfermions,spins}, where universal three-body correlations can be explored under a variety of interaction strengths, mass ratios, and confinement geometries.

Nowadays, ultracold atomic systems offer remarkable flexibility to manipulate both interactions and external confinement. Highly anisotropic trapping potentials~\cite{opticallattice} can
reshape atomic clouds into pancake-~\cite{BEC2D} or cigar-like~\cite{BEC1D01,BEC1D02} geometries, effectively realizing quasi-two- and one-dimensional
systems~\cite{reviewtrap}. Such configurations have enabled systematic investigations of few-body correlations under reduced dimensionality and have stimulated numerous theoretical studies of Efimov physics beyond three dimensions. In recent years, several theoretical approaches have been developed to explore the role of dimensionality in few-body systems, including mixed-dimensional setups where different particles experience distinct confinements~\cite{mixed1,mixed2}, effective field-theory descriptions of dimensional crossover~\cite{EFT01,EFT02}, and models based on squeezing the system with external potentials~\cite{squeezing01,squeezing02}. These efforts reflect the growing interest in understanding how universal three-body correlations evolve as the spatial dimensionality of the system is modified. A particularly efficient theoretical approach to model this deformation is to describe the system in terms of an effective noninteger spatial dimension $D$, allowing the dimensional crossover to be investigated continuously without introducing additional degrees of freedom~\cite{D1,D2,D3,D4}.

Within this framework, the Bethe--Peierls boundary condition leads to a
characteristic equation for the scaling exponents $s_n$~\cite{STMDdim}.
Because this equation is even in $s_n$, its roots occur in pairs
$\pm s_n$. In this work we follow
only the pair continuously connected to the Efimov channel and crossing
zero at the critical dimension, $D_c$. We denote its positive magnitudes in the
two regimes by $s_{0_E}>0$ and $s_{0_u}>0$. We therefore write this pair as
$\pm i s_{0_E}$ for $D>D_c$ and $\pm s_{0_u}$ for $D<D_c$; its two members
coalesce at $s_n=0$ when $D=D_c$. The initial subscript $0$
identifies the root pair followed here, while $E$ and $u$ distinguish the
Efimov and unatomic regimes. 
Physically, $s_n$ controls the short-distance hyperradial behavior of the
Faddeev components. An imaginary exponent produces a log-periodic wave
function and discrete scale invariance, whereas a real exponent produces a
homogeneous power law and continuous scale invariance.  In general, $s_{0_E}$ depends on both the spatial dimension and the mass ratio. For three identical bosons in $D=3$, $s_{0_E}\approx1.00624$. The binding energies of successive Efimov states,
ordered toward the three-body threshold, obey
\begin{equation}
    \frac{|E_3^{(j)}|}{|E_3^{(j+1)}|}
    =
    \exp\left(\frac{2\pi}{s_{0_E}}\right),
\end{equation}
where $j$ labels the Efimov states. Thus, $s_{0_E}$ fixes the geometric
spacing of the Efimov spectrum. As the effective dimension is reduced,
$s_{0_E}$ decreases and vanishes at $D=D_c$~\cite{STMDdim}; below $D_c$,
the relevant roots are real. The term
``unatomic''~\cite{unatomic1,unatomic2}, introduced as the atomic
counterpart of the unparticle concept proposed by Georgi~\cite{georgi} and
of the unnucleus discussed in nuclear systems by Hammer and
Son~\cite{unnucleus1}, denotes atomic few-body systems exhibiting such nontrivial
power-law behavior; see also
Refs.~\cite{unnucleus2,unnucleus3,unnucleus4,unnucleus5,unnucleus6} for
theoretical investigations of systems in which unnucleus-like behavior is expected to emerge. Here we use ``unatomic'' specifically for the
non-Efimovian, continuously scale-invariant branch of the zero-range
continuous-dimensional description; it is not intended as standard
terminology for every few-body system with scale-invariant behavior.

Previous studies have shown that the transition from the Efimov to the unatomic regime leaves clear signatures in observable quantities such as the large-momentum tails of the single-particle momentum distribution. In particular, the associated contact parameters, which encode short-range correlations, undergo pronounced variations as the system is driven toward the critical dimension, revealing the progressive suppression of Efimov correlations and the emergence of continuous scale invariance~\cite{unatomic1,unatomic2}.

However, the characterization of the momentum distribution at the critical dimension remains incomplete within the framework established so far. In previous analyses, the large-momentum tail at $D_c$ was found to be dominated by the two-body contact, while the three-body contribution becomes formally divergent, preventing a consistent description of the asymptotic behavior in terms of finite three-body correlations. From an experimental perspective, this feature would hinder the extraction of the three-body contribution from the momentum tail precisely at the transition point between discrete and continuous scale invariance. 

In this work we show that the three-body sector admits a finite and well-defined contribution at the critical dimension, leading to a consistent characterization of the momentum distribution in this regime. We demonstrate that the asymptotic tail contains additional logarithmic structures, including a squared-logarithmic term characterized by the coefficient $C_3''$ when mass imbalance is present. This reveals the emergence of an intermediate scaling structure at the transition point, characterized by logarithmic behavior that is neither log-periodic, as in the Efimov regime, nor a simple power law as in the unatomic regime. In addition, we further investigate the dimensional interval separating the Efimov and unatomic phases, where two- and three-body correlations coexist with comparable weight, combining structural elements of both scaling regimes: the two-body contact, present in the Efimov phase, together with a three-body contribution governed by the power-law scaling typical of the unatomic regime. In our previous work, we argued that this regime shrinks as the mass imbalance of the trimers increases. In the present work, we complete this picture by investigating the universal momentum distribution throughout this critical region and identifying its microscopic signatures.

The article is organized as follows. In Sec.~\ref{section2}, we analyze the emergence of the intermediate scaling region (ISR) between the Efimov and unatomic regimes, emphasizing its dependence on dimensionality and mass imbalance. In Sec.~\ref{section3}, we investigate the momentum-space structure at the Efimov–unatomic transition. We first examine the asymptotic behavior of the spectator function at the critical dimension, highlighting the emergence of a distinct logarithmic scaling. We then construct the single-particle momentum distribution and derive its asymptotic expansion, identifying the leading and subleading contributions. Finally, we extract the two-body contact and the three-body tail coefficients and discuss their dependence on the mass ratio, including the additional quadratic-logarithmic contribution required to characterize the momentum tail at the transition.

Technical details are presented in the appendices: Appendix~\ref{appA} reviews the three-body problem in position and momentum space, Appendix~\ref{appB} provides the derivation of the leading and next-to-leading contributions to the momentum distribution, and Appendix~\ref{appC} collects the two-body contact and three-body tail coefficients, completing the characterization of the Efimov–unatomic transition across different mass ratios.

\section{THE EMERGENCE OF THE INTERMEDIATE SCALING REGIME}
\label{section2}

Recent studies based on the continuous dimensional approach have shown that a narrow interval emerges between the critical dimension $D_c$, where $s_{0_E}$ vanishes, and a lower dimension $\overline{D}$, below which the asymptotic behavior of the momentum distribution is entirely governed by a three-body contribution exhibiting power-law scaling~\cite{unatomic1,unatomic2}. Although this intermediate region has been previously identified, its underlying structure and physical properties have not been systematically characterized. In particular, the dependence of its extent on the mass imbalance of the three-body system remains unexplored. In this section, we present an analysis of the relevant scaling-exponent pair across the full dimensional interval $2<D\leq3$, which allows us to identify and delimit the intermediate scaling regime. Furthermore, we quantify the width of this regime and demonstrate that it is strongly controlled by the mass ratio. To establish a direct connection with experimentally realizable systems, we map the effective dimension onto the aspect ratio of an external harmonic confinement, showing that accessing this regime requires fine control of the trapping geometry.

\begin{center}
\begin{figure}[h!]
{\includegraphics[width=8.6cm]{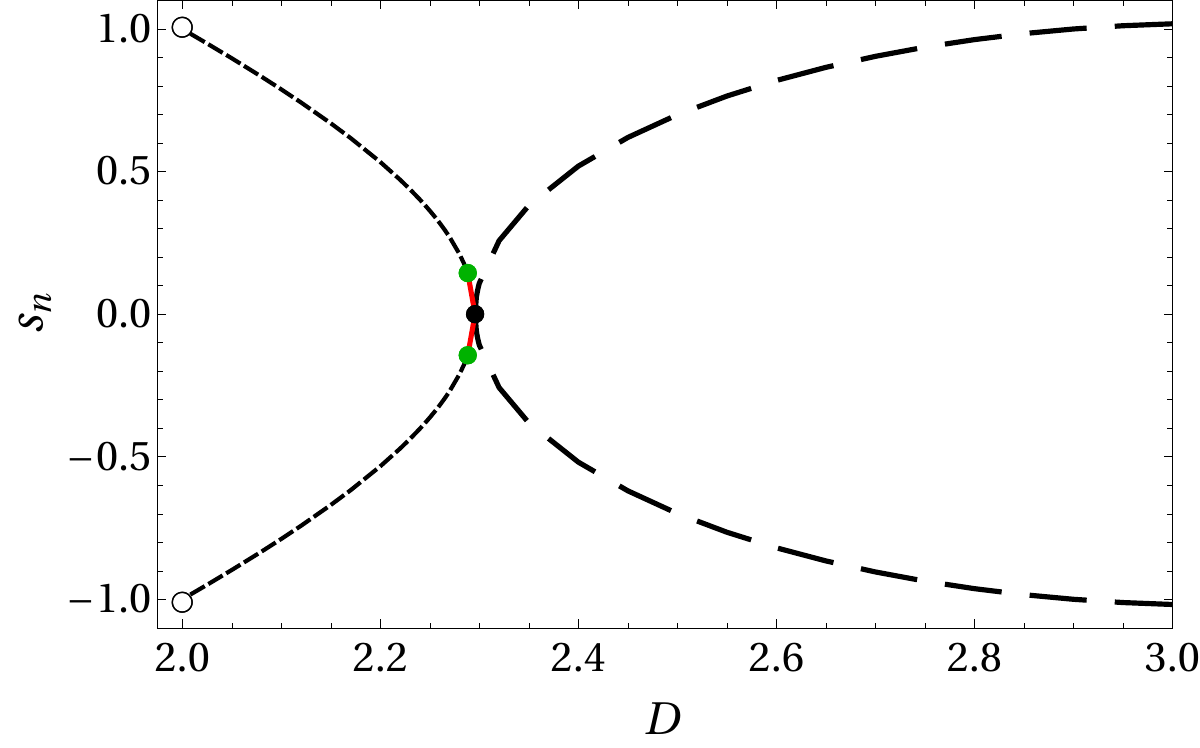}} 
{\includegraphics[width=8.6cm]{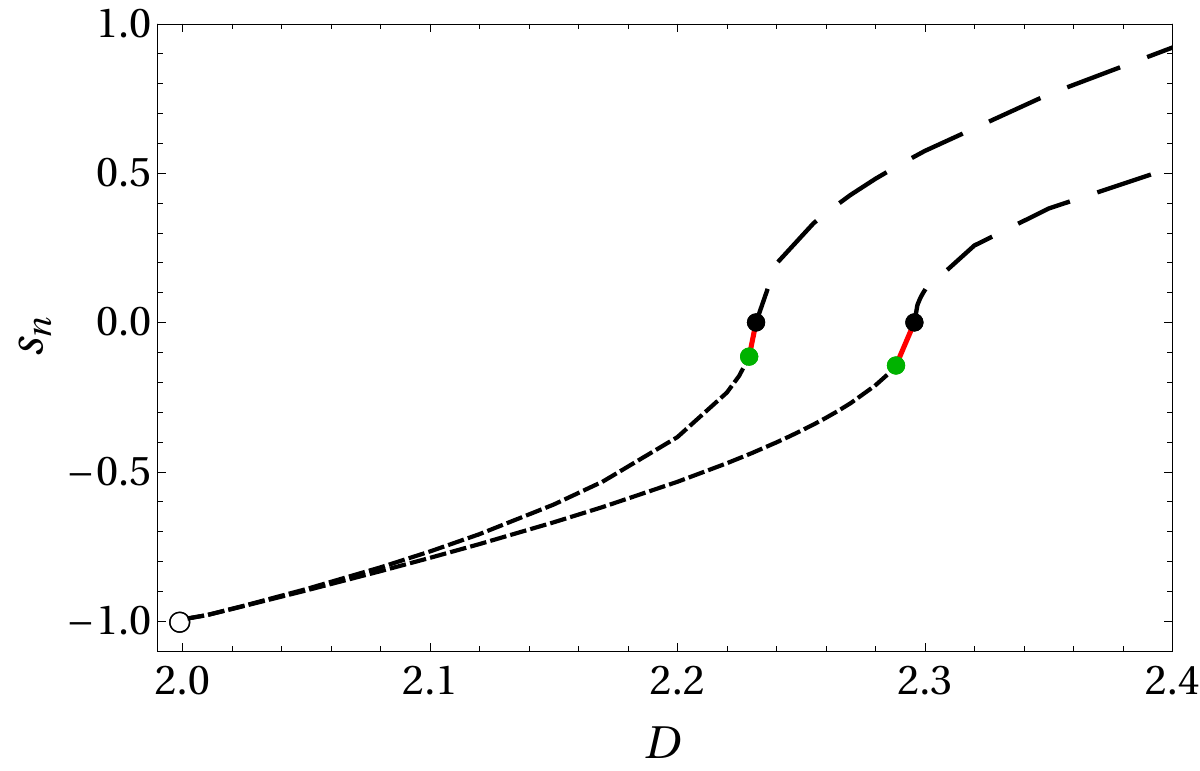}}
\caption{Scaling exponent $s_n$ as a function of the spatial dimension $D$ for resonant mass-imbalanced three-body systems. \textit{Upper panel}: Full behavior for the $^{23}$Na$_2$ – $^{40}$K system. The ordinate $s_n$ denotes the nonzero component of the exponent: the dashed curves show $\operatorname{Im}(s_n)=\pm s_{0_E}$ for $D>D_c$, and the dotted curves show $\operatorname{Re}(s_n)=\pm s_{0_u}$ for $D<D_c$. The positive quantities $s_{0_E}$ and $s_{0_u}$ are the magnitudes of the same root pair on the two sides of the transition; the roots coalesce at $s_n=0$ when $D=D_c$ (black point). The green point marks $\overline{D}$, below which the asymptotic behavior of the momentum distribution is entirely governed by a three-body contribution
exhibiting power-law scaling, characteristic of the unatomic regime. The red segment highlights the intermediate scale invariant region between $D_c$ and $\overline{D}$. The open circle indicates the limit $D \rightarrow 2+$. \textit{Lower panel}: Zoom of the transition region for two mass configurations, $^{23}$Na$_2$ – $^{40}$K and $^{133}$Cs$_2$ – $^{6}$Li. The displayed continuous representative uses the roots $+i s_{0_E}$ above $D_c$ and $-s_{0_u}$ below $D_c$; accordingly, its ordinate is $+s_{0_E}$ and $-s_{0_u}$, respectively. This sign choice is only a plotting convention. The curve with smaller $D_c$ (leftmost) corresponds to the more mass-imbalanced system. The separation between $D_c$ and $\overline{D}$ is seen to depend sensitively on the mass ratio.}
\label{fig1}
\end{figure}
\end{center}

Figure~\ref{fig1} shows the behavior of the scaling exponent $s_n$, obtained from the characteristic equation associated with three resonantly interacting atoms, as a function of the spatial dimension $D$. The details of the calculation leading to the scaling exponents are presented in Appendix~\ref{appA}. In the upper panel results are presented for the experimentally relevant mixture $^{23}$Na$_2$ – $^{40}$K~\cite{NaK}. In the interval $D_c<D\leq3$, the characteristic equation admits an infinite set of real solutions together with a single pair of imaginary ones, $\pm i s_{0_E}$. Here $n$ labels distinct root pairs, while $s_{0_E}>0$ is the magnitude of the pair followed in this work. The latter defines the attractive channel responsible for the Efimov effect and leads to discrete scale invariance. As the dimension is reduced, $s_{0_E}$ continuously approaches zero, and the two roots coalesce at $s_n=0$ when $D=D_c$, as indicated by the black point, signaling the disappearance of Efimov physics.

Below the critical dimension, $2<D<D_c$, the characteristic equation again yields an infinite set of real solutions. The continuation of the Efimov pair is $\pm s_{0_u}$, where $s_{0_u}>0$. Only this pair is relevant for describing the dimensional evolution of three-body states considered in this work. Imposing consistency with the Skorniakov–Ter-Martirosyan (STM) integral equation in noninteger dimensions~\cite{STMDdim} restricts the physically acceptable solutions to those that generate finite and normalizable spectator functions, constraining the scaling parameter to the interval $-1<s_n<1$, equivalently $0<s_{0_u}<1$ for its magnitude. The green point in Fig.~\ref{fig1} marks the dimension $\overline{D}$, which separates two qualitatively distinct regimes. For $D<\overline{D}$, the asymptotic momentum distribution is fully dominated by a three-body contribution with power-law scaling, characteristic of the unatomic regime. At this point, the two-body contribution becomes ultraviolet divergent and an exchange between leading and subleading terms in the asymptotic expansion takes place, as discussed in previous works~\cite{unatomic1,unatomic2}. 

Between $D_c$ and $\overline{D}$, we identify a narrow interval (red segment) defining an Intermediate Scaling Regime (ISR), where the system cannot be classified within either the Efimov or unatomic limits. Although narrow, this region anticipates a nontrivial restructuring of correlations, as neither log-periodic oscillations nor simple power-law scaling fully describe the system.

In the lower panel of Fig.~\ref{fig1}, we present a zoom of this transition region. Taking advantage of the symmetry of the spectator function under $s_n\rightarrow-s_n$, and consistently with our previous work, we display $s_n=+i s_{0_E}$ in the Efimov regime and $s_n=-s_{0_u}$ below $D_c$. The plotted ordinate is therefore $+s_{0_E}$ above $D_c$ and $-s_{0_u}$ below it, making the curve continuous through zero. This selection is needed only to display one branch: the two signs give the same physical solution and both are retained in the formulas. The results are shown for two experimentally relevant mass-imbalanced mixtures, $^{23}$Na$_2$ – $^{40}$K and $^{133}$Cs$_2$ – $^{6}$Li~\cite{CsLi}, illustrating how the transition between scaling behaviors is strongly influenced by mass imbalance. The curve with the smaller critical dimension corresponds to the more imbalanced system, consistent with the enhanced Efimov attraction in heavy–heavy–light configurations. In particular, the separation between $D_c$ and $\overline{D}$, and therefore the size of the ISR, is clearly mass dependent.

\begin{center}
\begin{figure}[h]
\includegraphics[width=8.7cm]{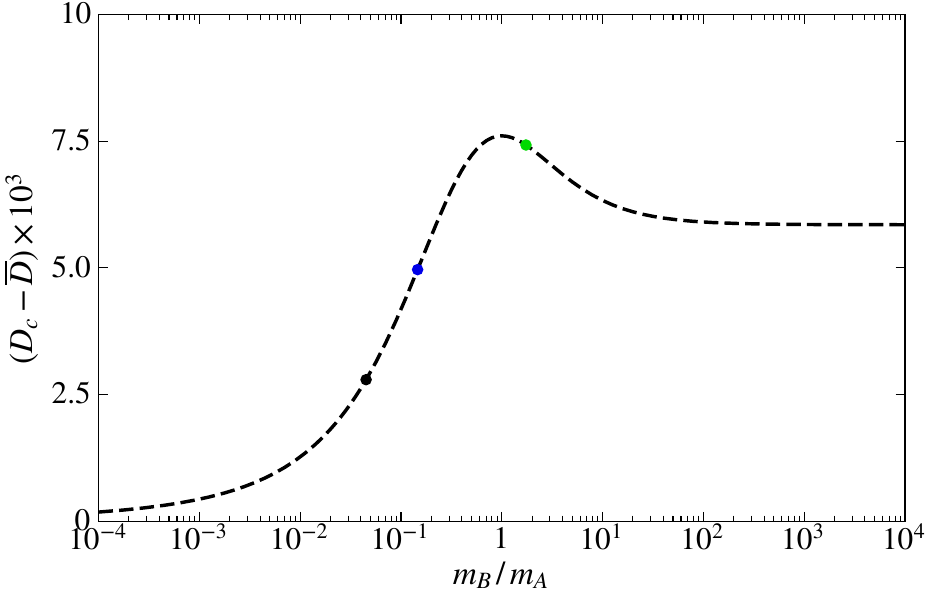}
\caption{Difference $D_c-\overline{D}$, which defines the width of the Intermediate Scaling Regime (ISR), as a function of the mass ratio $m_{B}/m_{A}$. The ISR corresponds to the interval in dimension where the system is not described by either Efimov (discrete scaling) or unatomic (continuous scaling) regimes. The highlighted points indicate experimentally relevant mixtures: $^{133}$Cs$_2$ – $^{6}$Li, $^{41}$K$_{2}$ – $^{6}$Li, and $^{23}$Na$_{2}$ – $^{40}$K.}
\label{fig2}
\end{figure}
\end{center}

To quantify this effect, in Fig.~\ref{fig2} we present the difference $D_c - \overline{D}$ as a function of the mass ratio $m_B/m_A$. This quantity provides a direct measure of the width of the ISR. The highlighted points correspond to mass-imbalanced systems, namely $^{133}$Cs$_{2}$ – $^{6}$Li, $^{41}$K$_2$ – $^{6}$Li, and $^{23}$Na$_2$ – $^{40}$K. We observe that the ISR becomes significantly narrower as the mass imbalance increases, indicating that the transition between discrete and continuous scaling symmetries becomes sharply localized in dimension for heavy–heavy–light systems. Conversely, more balanced systems exhibit a broader intermediate region.

A direct connection to experimental control parameters can be established by relating the effective dimension to the aspect ratio of the confining trap. Following previous works~\cite{connection}, this mapping can be expressed as
\begin{equation}
    \frac{3(D-2)}{(3-D)(D-1)} = \frac{b_{ho}^2}{r_{2D}^2},
\end{equation}
where $b_{ho}$ is the harmonic oscillator length and $r_{2D}$ is the root-mean-square radius of the three-body system in two dimensions. This relation, which is largely independent of the details of the interaction, allows one to translate variations in $D$ into experimentally tunable confinement parameters.

Using this mapping, we find that the variation in the aspect ratio required to span the interval between $D_c$ and $\overline{D}$ is extremely small. For the strongly imbalanced system $^{133}$Cs$_2$ – $^{6}$Li, the relative change in $b_{ho}/r_{2D}$ between these two points is below $1\%$, while for the more balanced $^{23}$Na$_2$ – $^{40}$K mixture it is of the order of $1.5\%$. This indicates that probing the ISR experimentally would require a high degree of control over the trap geometry.

The qualitative change in the nature of the scaling exponent across $D_c$, from imaginary to real through the marginal value $s_n=0$, anticipates a restructuring of three-body correlations at the transition. As will be shown in the following sections, this transition leaves clear signatures in the spectator function, the momentum distribution, and the associated two-body contact and three-body tail coefficients, providing a direct link between the underlying scaling structure and experimentally measurable observables.

\section{MOMENTUM-SPACE STRUCTURE AT THE EFIMOV-UNATOMIC TRANSITION}
\label{section3}

Having established the emergence of a narrow intermediate region between the Efimov and unatomic regimes, we now turn to the characterization of the momentum-space structure at the transition. In previous analyses, the behavior of the momentum distribution at the critical dimension $D_c$ suggested that three-body contributions become ill-defined~\cite{unatomic1,unatomic2}, preventing a consistent description of the asymptotic tail solely in terms of the known two-body contact and three-body tail coefficients. In particular, the extracted three-body contributions appear to diverge or become unreliable, allowing the misleading impression that the momentum tail at $D_c$ is entirely dominated by two-body correlations. This limitation poses a direct challenge for the interpretation of experiments probing the high-momentum distribution during dimensional crossover.

In this section, we show that the asymptotic behavior at $D_c$ is governed by a distinct scaling structure, characterized by a logarithmic dependence that differs fundamentally from both the log-periodic oscillations of the Efimov regime and the power-law scaling of the unatomic region. This observation indicates that the transition point cannot be described within the standard scaling frameworks and requires a refined treatment of the underlying three-body correlations. To elucidate the origin and implications of this behavior, we first analyze the asymptotic structure of the spectator function at the critical dimension, then construct the corresponding momentum distribution, and finally extract the associated two-body contact and three-body tail coefficients.

\subsection{Asymptotic spectator function at the critical dimension}
\label{subsection3a}

To identify the origin of the modified scaling behavior at the transition, we analyze the spectator function at the critical dimension $D_c$. Rather than repeating the full derivation presented in previous works, we refer the reader to Appendix~\ref{appA}, where the relevant steps are summarized. Here, we focus on the asymptotic structure that governs the large-momentum behavior. We start by obtaining the large-momentum behavior, $q_i^{\prime}\gg\kappa_0$, of the regular spectator function, Eq.~\eqref{regularspec}
 \begin{eqnarray}
&&\chi^{(i)}(q'_i)\underset{q_i^{\prime}  \gg \kappa_0}{=} \mathcal{C}^{(i)}\mathfrak{F}_{(D,s_n)}
\kappa_{0}^{1-D} \left(\frac{q_i^{\prime }}{\sqrt{2}\kappa_0}\right)^{1-D}\nonumber \\
&&\times\left[ \mathcal{G}_{(+ s_n)} \left(\frac{q_i^{\prime }}{\sqrt{2}\kappa_0}\right)^{s_n} +   \mathcal{G}_{(- s_n)} \left(\frac{q_i^{\prime }}{\sqrt{2}\kappa_0}\right)^{-s_n} \right],\hspace{1.1cm} 
 \label{asympespec}
\end{eqnarray} 
where
 \begin{eqnarray}
 \mathcal{G}_{(\pm s_n)}  
=\frac{\Gamma(\pm s_n)}{\Gamma\left[(D-1\pm s_n)/2\right]\Gamma\left[(1\pm s_n)/2\right]},
 \label{eq:G}
\end{eqnarray}
and 
\begin{eqnarray}
\mathfrak{F}_{(D,s_n)}& \equiv & \frac{2^{1+D/2} \pi^{D-1}\  \Gamma \big[ \mathcal{F}_{(D, s_n)}^{+}\big]\,
 \Gamma \big[ \mathcal{F}_{(D,s_n)}^{-}\big] }{(2-D)}\nonumber \\
 &\times&\cos \left[\frac{\pi}{2}(D-s_n)  \right]\csc \left( \frac{\pi}{2} s_n\right).
 \label{DefF}
\end{eqnarray}
For completeness, we also show the low-momentum region $q_i^{\prime} \ll \kappa_0$
\begin{eqnarray}
&&\chi^{(i)}(q'_i)\underset{q_i^{\prime}  \ll \kappa_0}{=} \mathcal{C}^{(i)}\mathfrak{F}_{(D,s_n)}
\kappa_{0}^{1-D} \frac{1}{\Gamma \left(D/2 \right)},
 \label{asympespeclowq}
\end{eqnarray} 
which will illustrate the smooth connection between the infrared behavior and the logarithmic asymptotics.

At $D_c$, the equation~\eqref{asympespec} leads to a logarithmic dependence on momentum
 \begin{eqnarray}
 \chi^{(i)}(q'_i)&\underset{q_i^{\prime}  \gg \kappa_0}{=} &\mathcal{C}^{(i)}\mathfrak{F}_{(D_c,s_n)}\frac{-2\ \kappa_0^{1-D_c}}{\sqrt{\pi}\Gamma\left[(D_c-1)/2\right]}\left(\frac{q_i^{\prime }}{\sqrt{2}\kappa_0}\right)^{1-D_c}\nonumber \\
&\times&  \ln\left[ \frac{2}{\sqrt{\exp\left[H_{(D_c-3)/2}\right]}} \frac{q_i^{\prime }}{\sqrt{2}\kappa_0}\right],\ \ \ \ \ \ 
 \label{asympspecdc}
\end{eqnarray}
where $H_n$ gives the $n^{\text{th}}$ Harmonic Number. The solution reflects the marginal nature of the scaling exponent at the transition. This behavior contrasts with both the oscillatory structure found in the Efimov regime and the power-law scaling characteristic of the unatomic region.

In Eq.~\eqref{asympspecdc}, $s_n\to0$ refers specifically to the root pair followed here approaching the critical point from either side of $D_c$. To make this limiting procedure explicit, we note that
\begin{eqnarray}
  \mathfrak{F}_{(D,s_n)}\big|_{s_n  \rightarrow 0} &=& -\frac{2^{2+D/2} \pi ^{D-2} \cos \left(\pi D /2\right) }{(D-2)\Gamma \left[(D-1)/2\right]^{-2} } \frac{1}{s_n}\nonumber \\
  &+& \mathcal{O}\left(s_n^{0}\right).
  \label{Fconst}
\end{eqnarray}
This quantity exhibits a singular behavior in the lowest-channel limit $s_n \rightarrow 0$, diverging as $1/s_n$. At first sight, this divergence might suggest that the solution becomes ill-defined at the critical dimension. However, no physical inconsistency arises. The regularized hypergeometric function in Eq.~\eqref{regularspec} is regular at ($s_n\rightarrow0$), while $\mathfrak{F}_{(D,s_n)}$ appears as a common multiplicative factor in all Faddeev components. Their relative weights are fixed by the coupled STM equations, whereas their common overall normalization remains arbitrary. We fix this remaining freedom by imposing unit normalization of the momentum distribution. Since $\mathfrak{F}_{(D,s_n)}\sim 1/s_n$ as $s_n\to0$, the normalization factor scale proportionally to $|s_n|$ compensating the divergence. Thus, the product of the normalization factor and $\mathfrak{F}_{(D,s_n)}$ remains finite in the limit ($s_n\to0$). Consequently, the normalized spectator functions, the momentum distribution, the two-body contact, and the three-body tail coefficients are finite at ($D_c$).

\begin{figure} 
{\includegraphics[width=8.55cm]{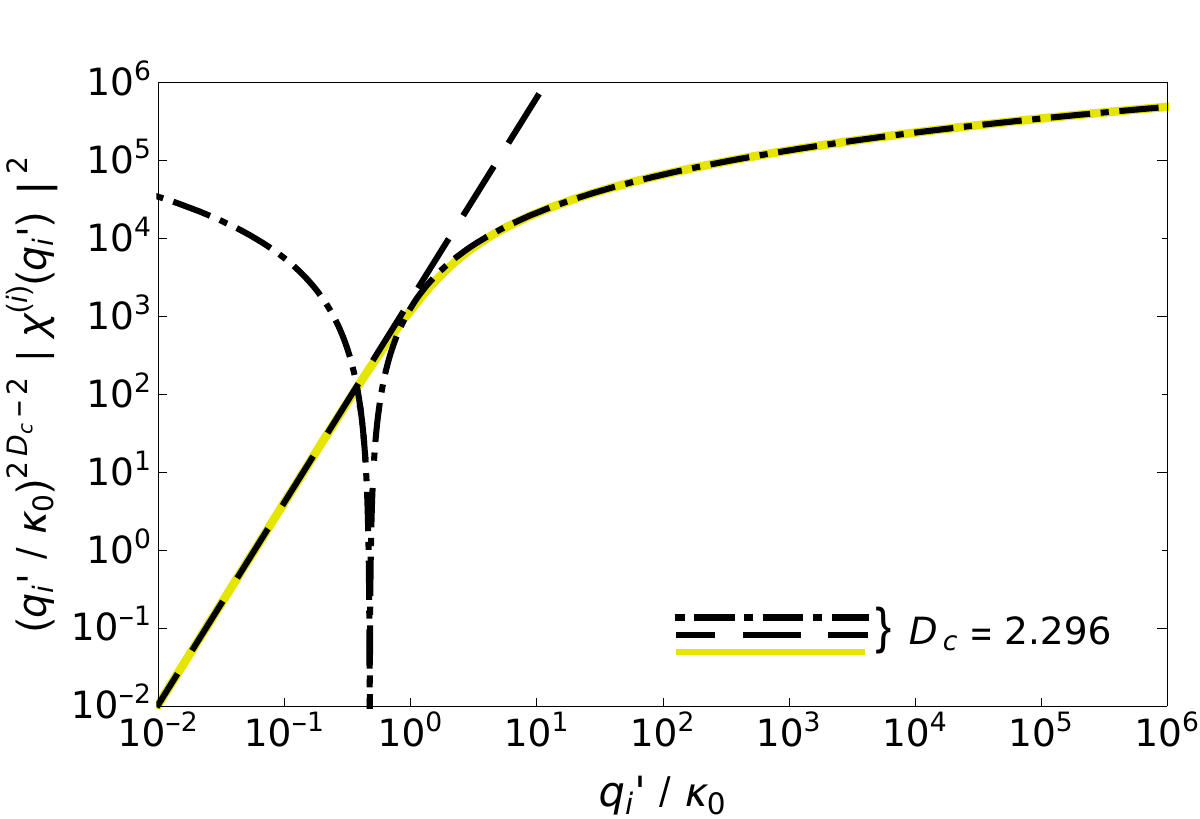}}
\caption{ Spectator function for the resonant system $^{23}$Na$_2$ - $^{40}$K at the critical dimension $D_c = 2.296$ ($b_{ho}/r_{2D} = 0.986$). The full solution, Eq.~\eqref{regularspec} (yellow), is compared with its high-momentum asymptotic form for $q_i^{\prime}\gg\kappa_0$ (black dot-dashed, Eq.~\eqref{asympspecdc}) and its low-momentum form for $q_i^{\prime}\ll\kappa_0$ (black dashed, Eq.~\eqref{asympespeclowq}), showing the rapid convergence of the logarithmic asymptotics at large momentum.}
\label{fig3}
\end{figure}

In Fig.~\ref{fig3}, we present the spectator function at $D_c$ for the mixture $^{23}$Na$_2$ – $^{40}$K together with its asymptotic form. For completeness, we also show the low-momentum region $q_{i}^{\prime} \ll \kappa_0$, which illustrates the smooth connection between the infrared behavior and the logarithmic asymptotics. A key feature observed in the figure is the rapid convergence of the asymptotic expression toward the full spectator function, indicating that the logarithmic form provides an accurate description over a broad momentum range. This property is particularly useful for the analysis of observable quantities, as it allows one to capture the relevant scaling behavior without relying on the full numerical solution.

\begin{center}
\begin{figure}[h]
{\includegraphics[width=8.2cm]{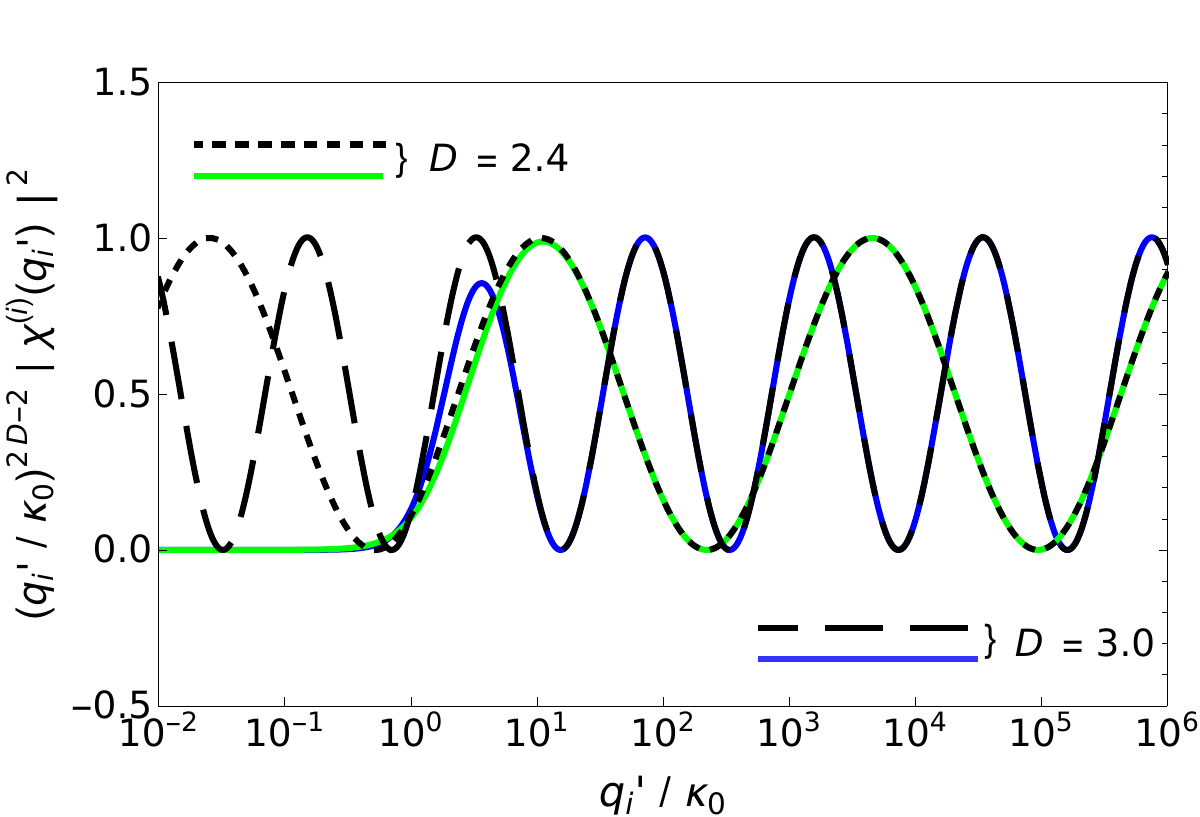}} 
{\includegraphics[width=8.2cm]{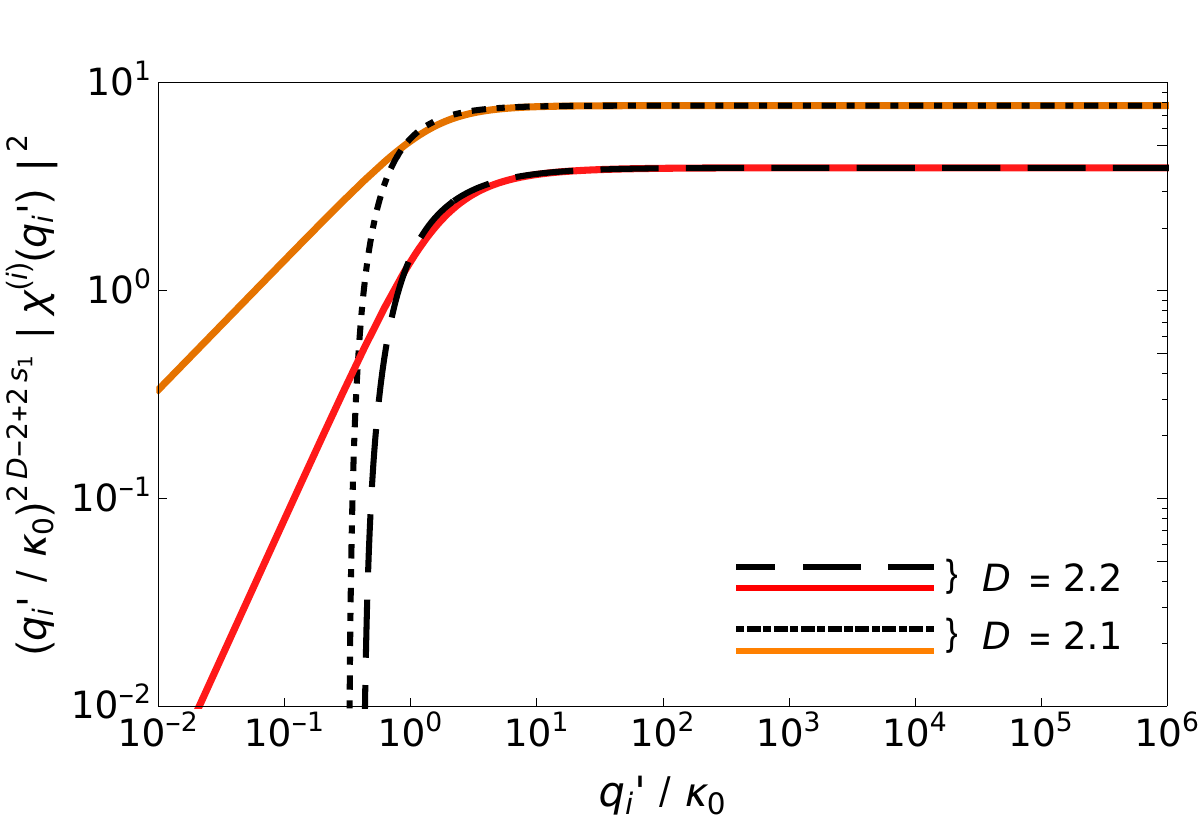}}
\caption{Spectator functions in momentum space for the resonant system $^{23}$Na$_{2}$ - $^{40}$K computed with Eq.~\eqref{regularspec} and compared with the zero-energy form in Eq.\eqref{asympespec}. \textit{Upper panel}: results at $D=3$ and $D=2.4$ ($b_{ho}/r_{2D}=1.195$). The finite-energy spectator functions (colored lines) are compared with their zero-energy forms (dashed and dotted black lines). \textit{Lower panel}: the same comparison in the unatomic regime.}
\label{fig4}
\end{figure}
\end{center}

To further emphasize the distinct nature of the critical dimension, Fig.~\ref{fig4} compares the spectator function in the Efimov and unatomic regimes for representative values of dimensions, where the spectator functions are shown with arbitrary normalization. In the Efimov region, shown in the upper panel, the spectator function exhibits the expected log-periodic oscillations associated with discrete scale invariance. Reducing the effective dimension decreases $s_{0_E}$, so that the distance between consecutive nodes of the wave-function increases~\cite{Nb_AAB_Ddim_Efimov}. In the unatomic regime~\cite{unatomic1,unatomic2}, lower panel, the spectator function follows a pure power-law behavior governed by a real scaling exponent. In both upper and lower panels, the spectator functions obtained for finite three-body binding energy (colored lines) are characterized by a damping in the low momentum region. In contrast, the critical dimension displays neither oscillatory nor power-law scaling, but instead a logarithmic dependence, signaling the emergence of a qualitatively different scaling structure. 

This comparison demonstrates that the critical dimension cannot be interpreted as a simple limiting case of either regime. Instead, it represents a distinct configuration in which the scaling properties of the three-body system are fundamentally altered, leading to observable consequences in the momentum distribution that will be explored in the following sections.

\subsection{Momentum-space structure at the critical dimension}
\label{subsection3b}

The momentum distributions in $D$-dimensions for the particles $A$ and $B$ are given, respectively, by
\begin{equation}
 n_A(q_A) = \int d^{D}p_A \ |\langle \textbf{q}_A \textbf{p}_A | \Psi \rangle |^{2} \,,
 \end{equation}
 and
\begin{equation}
\label{totdenB}
 n_B(q_B) = \int d^{D}p_B \ |\langle \textbf{q}_B \textbf{p}_B | \Psi \rangle |^{2} \,,
 \end{equation}
where $| \textbf{q}_A \textbf{p}_A \rangle$ and $| \textbf{q}_B \textbf{p}_B \rangle$ denote the corresponding Jacobi momentum basis states. The normalization is fixed by
\begin{equation} 
 \int d^Dq_B\, n_B(q_B)
 =1\,\,\text{and}\, \,
 \int d^Dq_A\, n_A(q_A)
 =1
 \, .
 \label{norm}
 \end{equation} 
 These distributions provide direct access to experimentally measurable quantities and encode the short-range correlations of the system. In particular, the leading large-momentum term is governed by the two-body
contact~\cite{Tan1,Tan2,Tan3} whereas the subleading terms encode three-body correlations~\cite{castindensity,yamashitadensity,musolino}. The physical status of the coefficients appearing
in the subleading tail at $D_c$ is specified below. Before addressing the asymptotic structure, it is instructive to analyze the regular momentum distribution obtained from the finite three-body energy spectator function~\eqref{regularspec}. This provides a global view of how the distribution evolves with the spatial dimension. 

\begin{figure}[t!]
\includegraphics[width=8.5cm]{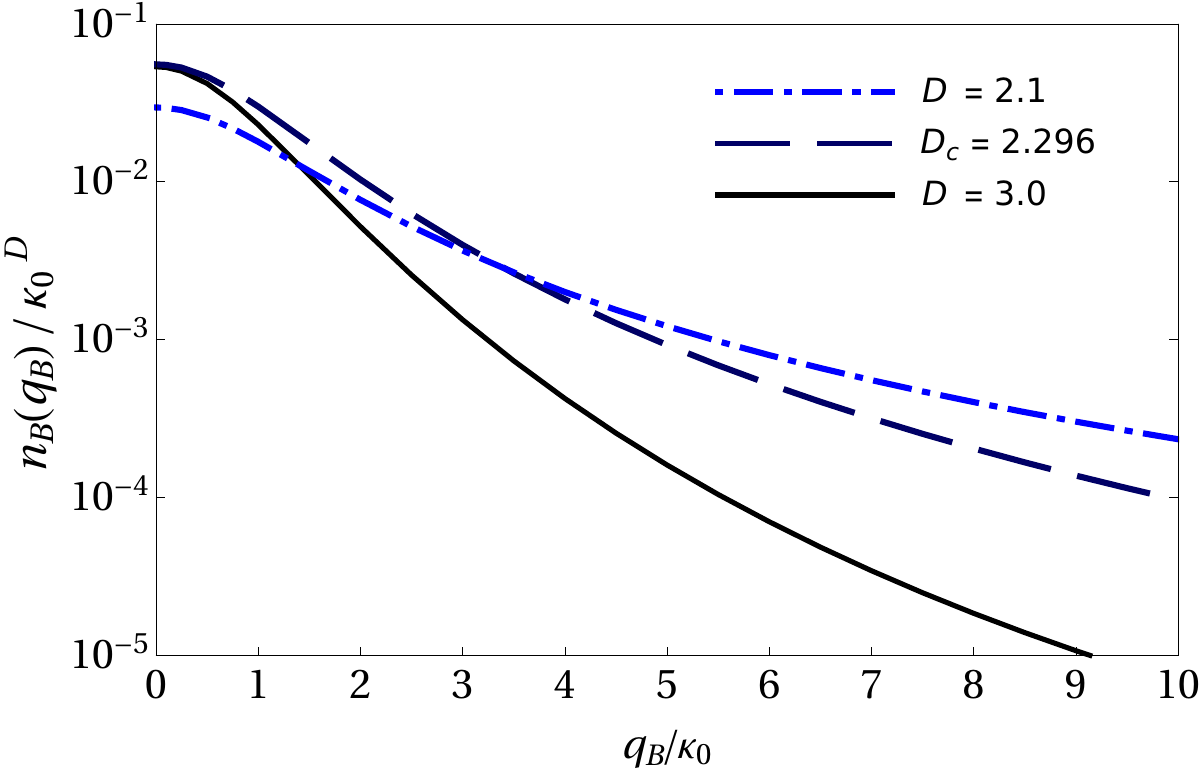}
\caption{Single particle momentum distribution of the resonant compound $^{23}$Na$_2$ - $^{40}$K for
different dimensions, namely, $D=3$, $D_c = 2.296$ (harmonic-trap length of $b_{ho}/r_{2D}=0.986$) and $D=2.1$. }
\label{fig5}
\end{figure}
 
The normalized momentum distribution $n_B(q_B)$ is shown in Fig.~\ref{fig5} for the resonant $^{23}$Na$_2$–$^{40}$K system at different dimensions, including $D=3.0$, $D_c$, and $D=2.1$. As the dimension is reduced, the system becomes progressively less localized in momentum space, leading to an enhancement of the high-momentum components. This behavior reflects a significant redistribution of correlations toward shorter distances as the system is squeezed. As will be demonstrated in the following, this enhancement of the large-momentum region directly impacts the two-body contact and three-body tail coefficients, which control the asymptotic behavior of the distribution.

Despite capturing the global features of the momentum distribution, Fig.~\ref{fig5} also illustrates that the damping of $n_B(q_B)$ at large momenta masks the detailed structure of the asymptotic tail. In particular, neither the log-periodic oscillations characteristic of the Efimov regime nor the power-law behavior of the unatomic regime can be unambiguously identified from the full distribution alone. This limitation highlights the necessity of isolating the leading and subleading contributions to the momentum distribution. The derivation of the asymptotic expansion, including the evaluation of the leading and next-to-leading terms, is presented in Appendix~\ref{appB}. In the following, we use these results to construct the asymptotic form of the momentum distribution at the critical dimension and to identify the corresponding two-body contact and three-body tail coefficients.

The asymptotic behavior of the momentum distribution can be obtained from the integral representations given in Appendix~\ref{appB}. Rather than repeating the full derivation, we directly present the resulting leading and subleading contributions at large momentum. At the critical dimension, the integral representations in Eqs.~\eqref{eq:n1desenvolved}, \eqref{eq:n2desenvolved}, \eqref{eq:n3desenvolved}, and \eqref{eq:n4desenvolved}, together with the asymptotic spectator function in Eq.~\eqref{asympspecdc}, give the following large-momentum expressions:
\begin{eqnarray}
 n_{1}(q_B) &=&  \frac{\lvert \mathcal{C}^{(B)} \rvert^{2}}{q_{B}^{D_c+2}}\mathcal{S}_{D_c} 
 \csc\left( \frac{D_c\pi}{2} \right)\left(2\mu_B\right)^{1+D_c/2}\nonumber \\
&\times&\frac{2-D_c}{\Gamma\left[(D_c-1)/2\right]^2}\nonumber \\
&\times&  \ln\left( \frac{2}{\sqrt{\exp\left[H_{(D_c-3)/2}\right]}} \frac{q_B}{\sqrt{2\mu_B}\kappa_0}\right) ^2,\nonumber \\
\label{n1dc}
\end{eqnarray}
\begin{eqnarray}
n_{2}(q_B) &-& \frac{C_2}{q_B^{4}}= \frac{|\mathcal{C}^{(A)}|^2}{q_B^{D_c+2}} \mathcal{S}_{D_c}  \frac{8\left(2\mu_A\right)^{D_c-1}}{\pi\Gamma\left[(D_c-1)/2\right]^2} \nonumber \\
&\times&\int^\infty_0 dq'_A \ q_A^{\prime 1-D_c}\left(\mathcal{H}(q'_A)-\frac{\mu_B ^2}{\mu_A^2}\right)\nonumber \\
&\times&    \ln\left( \frac{2}{\sqrt{\exp\left[H_{(D_c-3)/2}\right]}} \frac{q_B }{\sqrt{2\mu_A}\kappa_0}q'_A\right) ^2,\nonumber \\ 
\label{n2dc}
 \label{eq:B5}\end{eqnarray}  
\begin{eqnarray}
 n_{3}(q_B)&=& \frac{\mathcal{C}^{(B)^{*}} \mathcal{C}^{(A)}}{q_B^{D_c+2}}\frac{16\mathcal{S}_{D_c}}{\pi \Gamma\left[(D_c-1)/2\right]^2} \left(\sqrt{ \mu_B \mu_A}\right)^{1-D_c}\nonumber \\
 &\times&\ln\left( \frac{2}{\sqrt{\exp\left[H_{(D_c-3)/2}\right]}} \frac{q_B }{\sqrt{2\mu_B}\kappa_0}\right) \nonumber \\
 &\times&
 \int^\infty_0   dq'_A  \ 
 \mathcal{H}(q'_A)\nonumber \\
 &\times&\ln\left( \frac{2}{\sqrt{\exp\left[H_{(D_c-3)/2}\right]}} \frac{q_B }{\sqrt{2\mu_A}\kappa_0}q'_A\right)\,,\nonumber \\
 \label{n3dc}
 \end{eqnarray}
\begin{eqnarray}\label{n4efimov}
 n_{4}(q_B) &= &\frac{|\mathcal{C}^{(A)}|^2}{q_B^{D_c+2}} \frac{16 \pi^{(D_c-3)/2}}{\Gamma\left[(D_c-1)/2\right]^3}\nonumber   \\
 &\times&\int^\infty_0 dp'_B  \frac{p_B^{\prime D_c-1}  }
 { \left[ p_{B}^{\prime 2} + 1/2\mu_{B} \right]^{2} } \int_0^{\pi}d\theta (\sin\theta)^{D_c-2} \nonumber  \\
 &\times& \left( \frac{p'_{B_-} }{\sqrt{2\mu_A}} \right)^{1-D_c} \left( \frac{p'_{B_+} }{\sqrt{2\mu_A}} \right)^{1-D_c}\nonumber \\
 &\times&\ln\left[ \frac{2}{\sqrt{\exp\left[H_{(D_c-3)/2}\right]}} \frac{q_B\ p'_{B_-} }{\sqrt{2 \mu_A}\kappa_0}\right] \nonumber \\
 &\times&\ln\left[ \frac{2}{\sqrt{\exp\left[H_{(D_c-3)/2}\right]}} \frac{q_B\ p'_{B_+} }{\sqrt{2 \mu_A}\kappa_0}\right].\ \ \ \ \
 \label{n4dc}
 \end{eqnarray}

 Collecting all four terms, the leading and sub-leading
contributions in the asymptotic region of the momentum distribution is summarized as 
\begin{eqnarray}
\hspace{-0.2cm}n_B(q_B) \underset{q_B  \gg \kappa_0}{=}  \frac{C_{2}}{q_B^{4}}+\frac{C_{3}'}{q_B^{D_c+2}} +\frac{C_3}{q_B^{D_c+2}} \ln\left(\frac{q_B/\kappa_0}
{(4\mu_A \mu_B )^{1/4} }  \right)\hspace{-0.8cm}\nonumber \\
+ \frac{C_3''}{q_B^{D_c+2}} \left[ \ln\left(\frac{q_B/\kappa_0}
{(4\mu_A \mu_B )^{1/4} }  \right)\right]^2+\cdots ,\hspace{0.9cm}
\label{tailnbdc}
\end{eqnarray}
it reveals that, at the critical dimension, the asymptotic structure of the momentum distribution differs fundamentally from both the Efimov and unatomic regimes. Instead of the log-periodic oscillations characteristic of Efimov physics or the pure power-law scaling associated with the unatomic regime, the expansion exhibits a logarithmic hierarchy involving constant, linear, and quadratic logarithmic terms. This structure reflects the marginal nature of the scaling exponent at $D_c$, where the transition between discrete and continuous scale invariance takes place.

The
dependence on the external momentum of Eq.~\eqref{tailnbdc}  can nevertheless be
isolated by introducing a common reference scale. We adopt
\begin{equation}
q_\star
\equiv
\kappa_0(4\mu_A\mu_B)^{1/4}=\sqrt{k_Ak_B},
\label{qstar}
\end{equation}
with $k_i=\kappa_0\sqrt{2\mu_i}$ the binding-momentum scale associated with spectator channel $i$. Thus, $q_\star$ is the geometric mean of the two channel-dependent momentum scales and provides a symmetric convention for the logarithmic terms. With
this choice, the large-momentum tail can be written as
\begin{equation}
\begin{split}
n_B(q_B)
=
\frac{C_2}{q_B^4}
&+
\frac{1}{q_B^{D_c+2}}
\left[
C_3'
+
C_3\ln\left(\frac{q_B}{q_\star}\right)
\right. \\
&\left.
+
C_3''\ln^2\left(\frac{q_B}{q_\star}\right)
\right] +\cdots .
\end{split}
\end{equation}
The choice of $q_\star$ is conventional. To make the
corresponding scale dependence explicit, consider another
reference momentum
$\widetilde q_\star=e^\delta q_\star$. Requiring the
momentum distribution to remain unchanged gives
\begin{align}
\widetilde C_3'' &= C_3'', \\
\widetilde C_3   &= C_3+2\delta C_3'', \\
\widetilde C_3'  &= C_3'+\delta C_3+\delta^2 C_3'' .
\end{align}
Thus, for $C_3''\neq0$, $C_3'$ and $C_3$ separately
depend on the logarithmic scale convention, whereas
$C_3''$ is invariant. All numerical values of $C_3'$ and $C_3$ quoted
below refer to the convention defined in Eq.~\eqref{qstar}.

Equation~\eqref{tailnbdc} reveals that, at the critical dimension, the asymptotic structure of the momentum distribution differs fundamentally from both the Efimov and unatomic regimes. Instead of the log-periodic oscillations characteristic of Efimov physics or the pure power-law scaling associated with the unatomic regime, the expansion exhibits a logarithmic hierarchy involving constant, linear, and quadratic logarithmic terms. This structure reflects the marginal nature of the scaling exponent at $D_c$, where the transition between discrete and continuous scale invariance takes place.

In Eq.~\eqref{tailnbdc}, $C_2$, defined in Eq.~\eqref{c2}, is the two-body
contact, whereas $C_3'$, $C_3$, and $C_3''$ denote the
coefficients of the constant, linear-logarithmic, and
quadratic-logarithmic terms, respectively, in the subleading
momentum tail. These coefficients characterize the three-body
contribution to the high-momentum distribution and can, in
principle, be extracted from momentum-distribution measurements~\cite{fletcher0,wild,makotyan}.
Within the unit-normalized zero-range trimer considered here,
their values are fixed once the mass ratio and $\kappa_0$ are
specified. In particular, $C_3''$ is
defined operationally in the present work as the asymptotic
tail coefficient multiplying the quadratic-logarithmic term
in Eq.~\eqref{tailnbdc}. A central result of this
work is that this contribution is finite at $D_c$ for
mass-imbalanced systems, is absent in both the Efimov and
unatomic limits, and emerges specifically at the transition,
reflecting the distinct scaling structure of this regime.

This logarithmic hierarchy is directly connected to earlier
developments in the momentum tails of Efimov systems. For three
identical bosons in three dimensions, the subleading momentum
tail exhibits the characteristic log-periodic structure of
Efimov physics~\cite{castindensity}. For mass-imbalanced
systems, two coefficients, $C_3$ and $C_3'$, are required to
characterize the corresponding oscillatory
structure~\cite{yamashitadensity}. In the present context, the
emergence of $C_3''$ plays an analogous role, but is now
associated with the logarithmic structure at the transition
between scaling regimes.

\begin{figure}[t!]
\includegraphics[width=8.5cm]{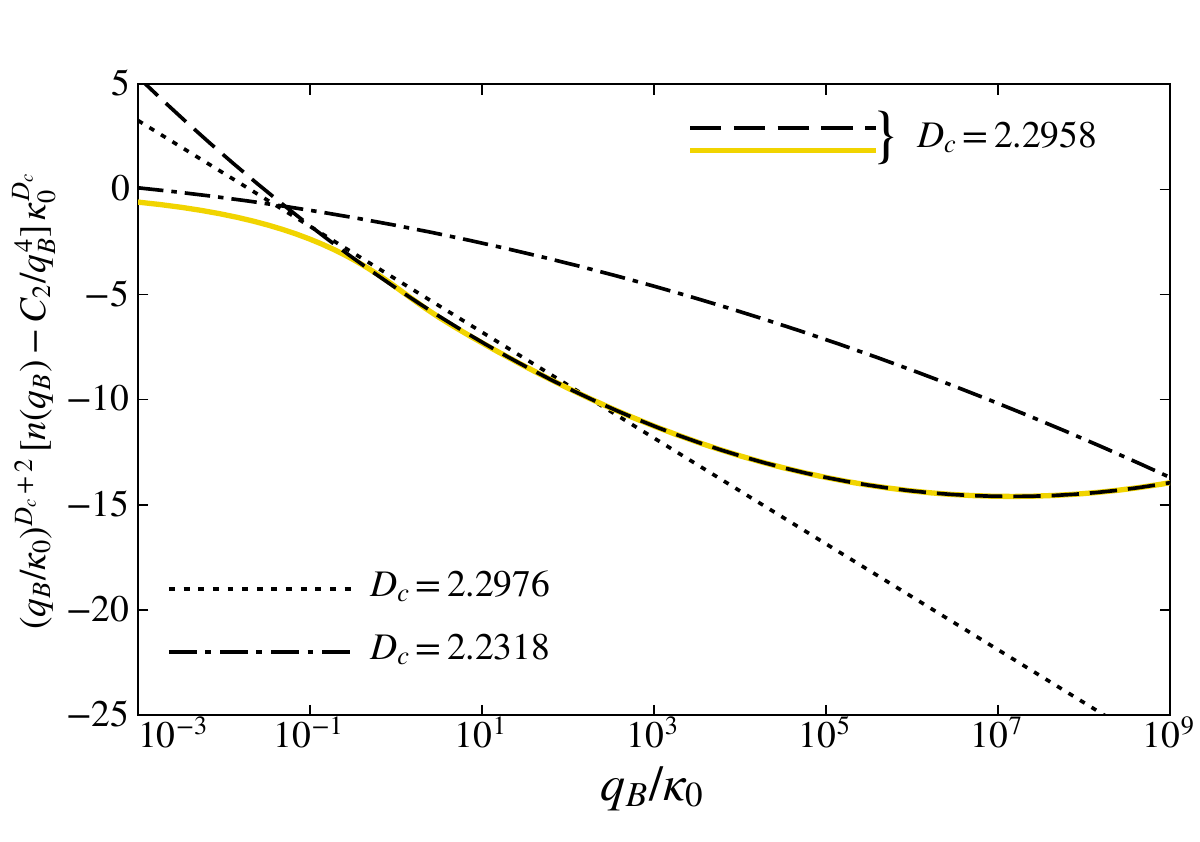}
\caption{Subtracted single-particle momentum distribution $\left[n_B(q_B)-C_2/q_B^{4}\right]q_B^{D_c+2}/\kappa_0^2$ at the critical dimension $D_c$ for different mass configurations. Results are shown for the resonant systems $^{23}$Na$_2$–$^{40}$K (yellow solid line: full numerical calculation; dashed black line: asymptotic expression) and $^{6}$Li–$^{133}$Cs$_2$ (dot-dashed black line: asymptotic result). The case of three identical particles is also included for comparison (dotted black line: asymptotic result). For $^{23}$Na$_2$–$^{40}$K, the visible change of slope corresponds to a minimum near $q_B=1.45\times10^7\kappa_0$. Equation~\eqref{tailnbdc} predicts the high-momentum zero of the retained three-body term at $q_{B_c}=7.18\times10^{15}\kappa_0$, outside the range shown. For $^{6}$Li–$^{133}$Cs$_2$, $C_3''<0$ and no high-momentum zero occurs, whereas for three identical particles $C_3''$ is consistent with zero.}
\label{fig6}
\end{figure}

In Fig.~\ref{fig6} we plot the subtracted momentum distribution for different mass configurations at the critical dimension. The results reveal a pronounced dependence of the high-momentum structure on the mass ratio. For the $^{23}$Na$_2$–$^{40}$K system, both the full numerical result (yellow line) and the asymptotic expression (dashed black line) are shown, displaying excellent agreement in the large-momentum region. In this case, $C_3''>0$, so that the asymptotic curve exhibits a clear change of slope, corresponding to a minimum at $q_B=1.45\times10^7\kappa_0$. Besides that, Eq.~\eqref{tailnbdc} predicts a high-momentum zero at $q_{B_c}=7.18\times10^{15}\kappa_0$. At this zero, the three-body term retained in Eq.~\eqref{tailnbdc} cancels, leaving the momentum distribution locally governed, at the order considered, by the two-body contact $C_2$. For more strongly mass-imbalanced systems, such as $^{6}$Li–$^{133}$Cs$_2$, we display only the asymptotic behavior (dot-dashed black line). In this configuration, no high-momentum zero occurs because $C_3''<0$. In contrast, the case of three identical particles (dotted black line) corresponds to the limiting situation where the logarithmic-squared contribution is equal to zero. In this case, the asymptotic structure is governed by the remaining logarithmic term. Therefore, the emergence of the $C_3''$ term and its mass dependence, which is discussed in detail below, provide a direct manifestation of the interplay between dimensionality and mass imbalance at the Efimov–unatomic transition. 

\begin{center}
\begin{figure}[h]
{\includegraphics[width=8.2cm]{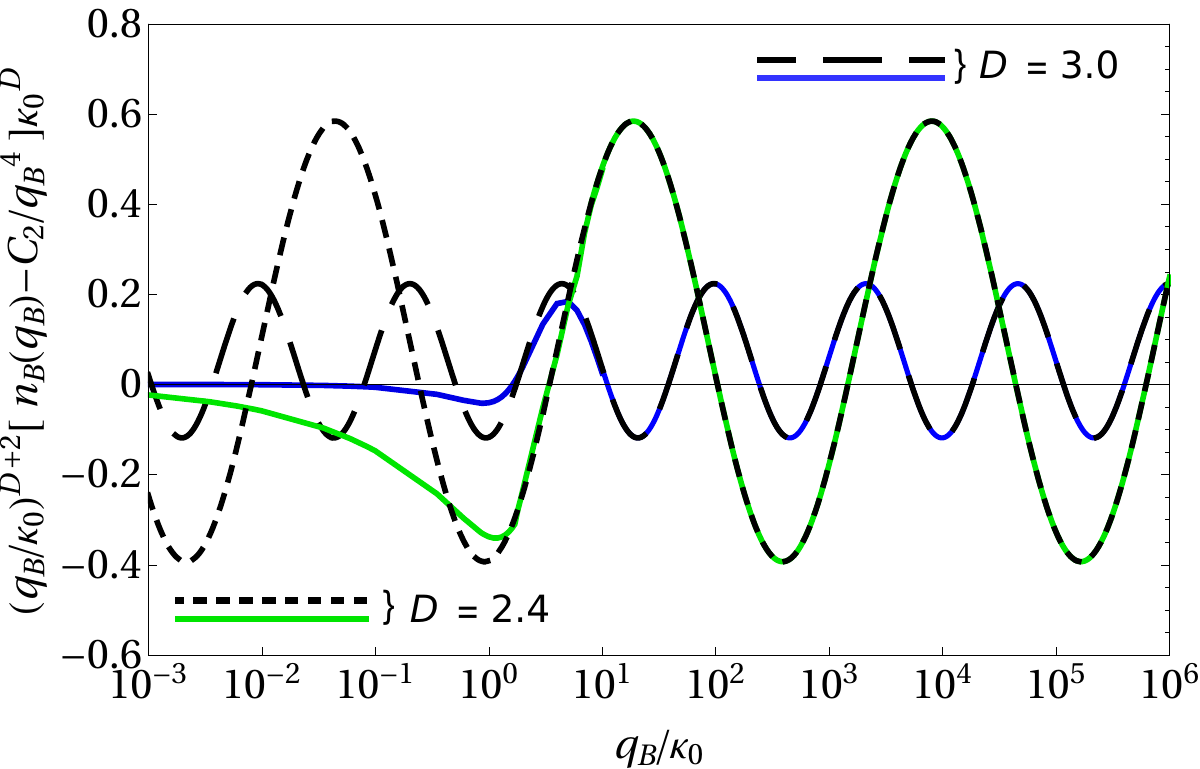}} 
{\includegraphics[width=8.2cm]{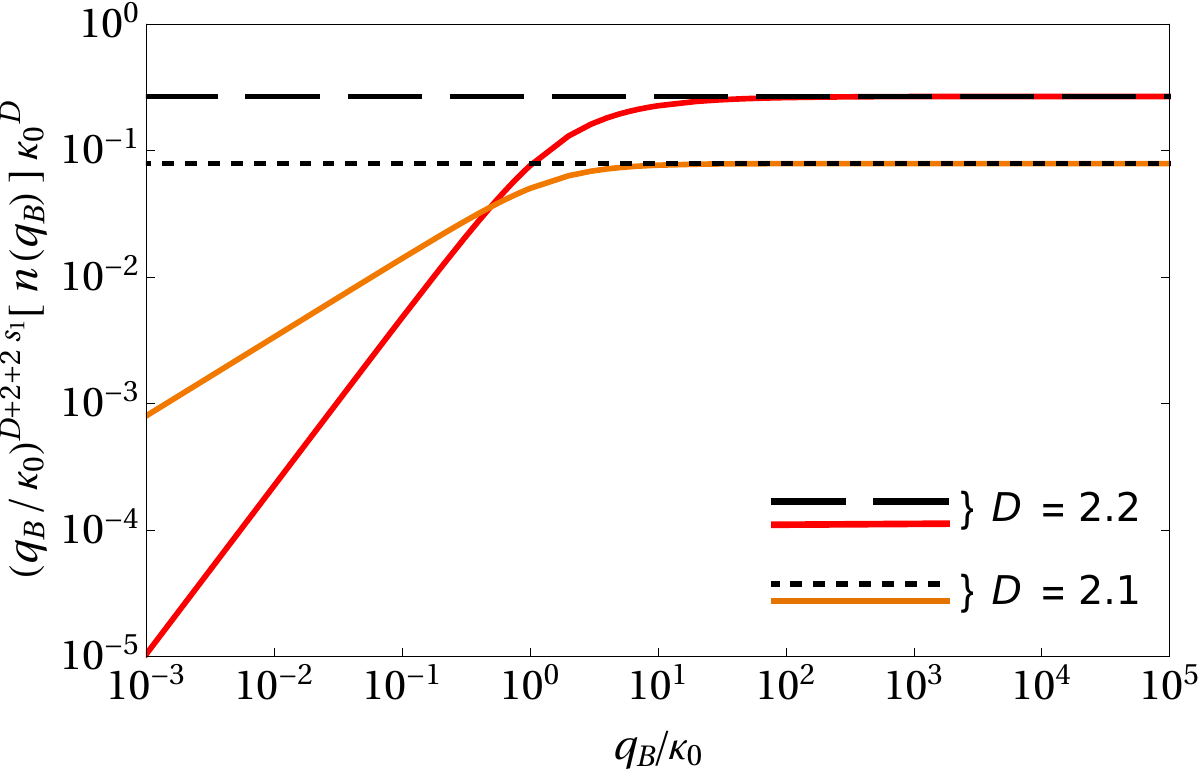}}
\caption{Subtracted single-particle momentum distribution for the resonant $^{23}$Na$_2$–$^{40}$K system in different dimensional regimes. \textit{Upper panel} (Efimov regime): Results for $D=3$ (blue solid line) and $D=2.4$ (green solid line), compared with the corresponding asymptotic expressions (black dashed and dotted lines, respectively). The log-periodic oscillations become progressively stretched as the dimension is reduced, reflecting the decrease of the Efimov scaling parameter $s_{0_E}$. \textit{Lower panel} (unatomic regime): Results for $D=2.2$ (red solid line) and $D=2.1$ (orange solid line), together with their asymptotic limits (black dashed and dotted lines). The momentum distribution exhibits a smooth power-law behavior, with decreasing amplitude as the dimension is lowered. The comparison highlights the qualitative difference between discrete scale invariance in the Efimov regime and continuous scale invariance in the unatomic regime.}
\label{fig7}
\end{figure}
\end{center}

To further elucidate the structural differences between the results at the critical dimension, Efimov and unatomic regimes, in Fig.~\ref{fig7} we present the subtracted momentum distribution for the resonant $^{23}$Na$_2$–$^{40}$K system in representative dimensions above and below the critical point, comparing the full numerical results with their corresponding asymptotic expressions. In the upper panel, corresponding to the Efimov regime, we show results for $D=3$ and $D=2.4$. The characteristic log-periodic oscillations are visible at large momenta. As the dimension is reduced, the separation between consecutive nodes increases, reflecting the decrease of the Efimov scaling parameter $s_{0_E}$. This behavior is consistent with the divergence of the log-periodic wavelength as $s_{0_E} \to 0$ when approaching the critical dimension. In addition, both the amplitude of the oscillations and their average increase as $D$ decreases, indicating an enhancement of the subleading three-body tail contributions. In the lower panel, corresponding to the unatomic regime, we present results for $D=2.2$ and $D=2.1$. In this case, the oscillatory structure is absent, and the asymptotic behavior is governed by a smooth power-law dependence. As the dimension is further reduced, the overall magnitude of the distribution decreases, signaling a suppression of three-body correlations. This trend reflects the reduction of the three-body tail coefficients in the unatomic regime, where continuous scale invariance replaces the discrete Efimov structure.

Taken together, these results demonstrate that the asymptotic momentum distribution undergoes a qualitative restructuring across the transition: from oscillatory log-periodic behavior in the Efimov regime to a monotonic power-law scaling in the unatomic regime. This sharp contrast further emphasizes that the critical dimension cannot be described as a simple interpolation between the two limits, but instead corresponds to a distinct regime with its own scaling structure, as discussed in the previous section.

\subsection{Two-body contact and three-body tail coefficients} \label{sec:contact} 

\begin{center}
\begin{figure}[h!]
\includegraphics[width=8.5cm]{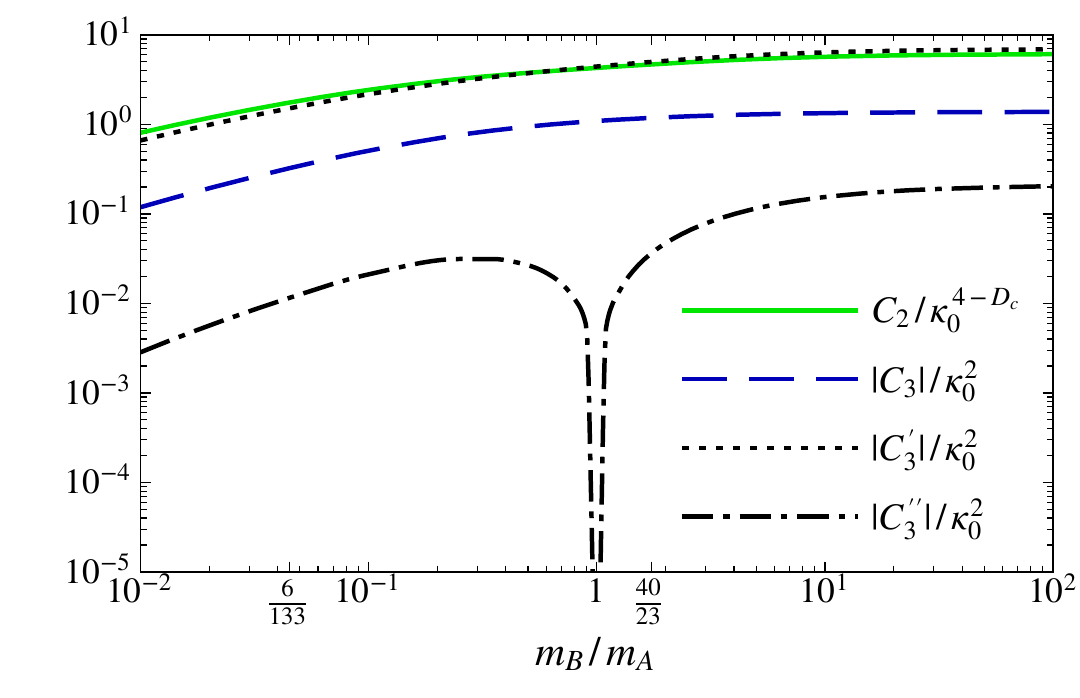}
\caption{Two-body contact and three-body tail coefficients at the critical dimension $D_c$ as functions of the mass ratio $m_B/m_A$. The plotted quantities correspond to $C_2$ (solid line), $|C_3|$ (dashed line), $|C_3'|$ (dotted line), and $|C_3''|$ (dot-dashed line). While $C_2$ remains positive and $C_3$, $C_3'$ are negative over the entire range, the coefficient $C_3''$ exhibits a sign change: it is negative for heavy-heavy-light configurations, vanishes for equal masses, and becomes positive in the heavy-light-light regime. This sign inversion plays a crucial role in determining the structure of the high-momentum tail at the critical dimension.}
\label{fig8}
\end{figure}
\end{center}

The asymptotic structure derived at the critical dimension allows for a direct extraction of the corresponding two-body contact $C_2$ and the three-body tail coefficients $C_3$, $C_3'$, and $C_3''$. In Fig.~\ref{fig8} we present the tail parameters at $D_c$ as functions of the mass ratio for resonant three-body systems. The quantities $C_2$, $C_3$, $C_3'$, and $C_3''$ are obtained from collecting the asymptotic form of the momentum distribution, Eq.~\eqref{tailnbdc}, and are shown in dimensionless units.

We observe that the two-body contact $C_2$ remains positive over the entire range of mass ratios, while both $C_3$ and $C_3'$ are negative. In contrast, the coefficient $C_3''$, associated with the logarithmic-squared contribution, exhibits a qualitative change of behavior as the mass imbalance is increased. In particular, $C_3''$ is negative for heavy-heavy-light (HHL) configurations, vanishes for three identical particles, and becomes positive in the heavy-light-light (HLL) regime. This sign inversion constitutes a central result of the present work, as it directly impacts the structure of the high-momentum tail.

As a consequence, in systems (HLL) where $C_3''>0$ , the three-body contributions can cancel at a finite momentum scale ($q_{Bc}$). The value of $q_{Bc}/\kappa_0$ can be extracted from
\begin{equation}
    \left[n(q_B)- C_2 / q_B^{4}\right] q_B^{D_c+2} = 0,
\end{equation}
in which the asymptotic behavior is effectively governed solely by the two-body contact $C_2$. Notably, this feature is absent in HHL systems, where $C_3''<0$, and therefore no real solution exists for such cancellation.
For mass configurations of the form Heavy-Light-Light, we find ($C_3'<0$), ($C_3<0$), and ($C_3''>0$). The three-body contribution is a quadratic function in $\ln(q_B/\kappa_0)$. Although this equation has two real roots in ($q_{Bc}$), one corresponds to ($q_{Bc}<\kappa_0$), and therefore lies outside the high-momentum regime. The physically relevant cancellation point is the unique root given by
\begin{equation}
    q_{Bc}=\kappa_0\exp\left[
\frac{-C_3+\sqrt{C_3^2-4C_3'C_3''}}
{2C_3''}
\right],
\end{equation}
which satisfies ($q_{Bc}>\kappa_0$). Thus, the sign change of ($C_3''$) with mass imbalance not only modifies the logarithmic structure of the critical tail, but can also produce a finite momentum scale at which the three-body contribution vanishes.

For a mass-imbalanced system of the form $^{133}$Cs-$^{6}$Li$_2$, the predicted zero occurs at  $q_{Bc}/\kappa_0 \sim 10^5$. It is worth mentioning that both the regular spectator-function calculation and momentum density are obtained within the zero-range theory. In a finite-range system, the momentum must also remain within the zero-range window, $q_B r_0 \ll 1$. The experimental accessibility requires $\kappa_0 r_0 \ll 10^{-5}$ and therefore depends on the particular interaction range, binding scale $\kappa_0(D)$, and confinement trajectory. The zero-range calculation applied in this work alone cannot establish whether this condition is satisfied for current experimental technologies. Nevertheless, the existence of this cancellation scale is a genuine feature of the universal zero-range description, and its experimental accessibility in realistic finite-range systems deserves detailed investigation in future works.

\section{SUMMARY AND CONCLUSIONS}
\label{section4}

In this work, we have investigated the dimensional evolution of a resonant mass-imbalanced three-body system by solving the Faddeev equations supplemented with the Bethe–Peierls boundary condition at infinite scattering length. By embedding the system in a continuous, noninteger spatial dimension $D$, we tracked within a unified framework the connection between distinct scaling regimes that are otherwise separated in fixed dimensions.

Our analysis shows that dimensionality acts as an effective control parameter driving the evolution from discrete to continuous scale invariance. In particular, the Efimov regime, characterized by log-periodic behavior and an imaginary scaling exponent, evolves continuously toward a regime where the scaling exponent becomes real and the system exhibits power-law behavior. This change occurs at the critical dimension $D_c$, where the Efimov parameter vanishes and the scaling structure is qualitatively modified.

A central result of this work is the characterization of the momentum-space structure at the critical dimension. The leading high-momentum term retains the universal $1/{q_B}^4$ scaling governed by the two-body contact $C_2$, while the subleading contribution contains constant, linear-logarithmic, and quadratic-logarithmic terms. The coefficient $C_3''$ of the quadratic-logarithmic term depends sensitively on the mass imbalance and changes sign across different mass configurations. Consequently, in HLL systems, the three-body term retained in Eq.~\eqref{tailnbdc} can vanish at a finite high-momentum scale. At the cancellation point, the momentum distribution is governed by the two-body-contact term only to the order retained in Eq.~\eqref{tailnbdc}. 

In addition, we have analyzed the narrow intermediate region $\overline{D}<D<D_c$, here identified as an intermediate scaling regime, where neither Efimov nor unatomic descriptions are sufficient. We show that its extent is strongly controlled by the mass ratio and becomes increasingly narrow for strongly mass-imbalanced systems. By relating the effective dimension to the aspect ratio of an external harmonic confinement, we demonstrate that accessing this regime requires fine control of the trapping geometry.

Finally, we have discussed how these results may extend beyond atomic systems. In nuclear halo systems, intrinsic physical scales such as finite effective range and centrifugal barriers can play a role analogous to external confinement, potentially driving the system toward the boundary between discrete and continuous scaling. From this perspective, the present framework establishes a bridge between ultracold atoms and nuclear few-body systems.

In summary, tuning dimensionality provides a powerful route to explore the interplay between few-body correlations and scale invariance. The emergence of a distinct logarithmic structure at the critical dimension, together with its signatures in momentum-space observables, opens new directions for experimental and theoretical investigations in universal few-body physics.


\section*{ACKNOWLEDGMENTS}
This work was partially supported by: Funda\c{c}\~ao de Amparo \`a Pesquisa do 
Estado de S\~ao Paulo (FAPESP) [grant numbers 2019/07767-1 (T.F.) 2018/25225-9 (G.K.) and 2023/08600-9 (R. M. F.)], Conselho Nacional de Desenvolvimento 
Cient\'{i}fico e Tecnol\'{o}gico (CNPq) [grant numbers 306834/2022-7 (T.F.), 151403/2025-2 (D. S. R.), 
302105/2022-0 (M.T.Y.) and 309262/2019-4 (G.K.)] and Coordenação de Aperfeiçoamento de Pessoal de Nível Superior (CAPES) Grant number
88887.928099/2023-00 (D.S.R).  This work is a part of the
project Instituto Nacional de  Ci\^{e}ncia e Tecnologia - F\'{\i}sica
Nuclear e Aplica\c{c}\~{o}es  Proc. No. 464898/2014-5.


\appendix

\section{The three-body problem in $D$-dimensions}
\label{appA}

\subsection{Position space}

In position space, we consider three different particles with masses $m_i$, $m_j$, $m_k$, and coordinates $\textbf{x}_{i}$, $\textbf{x}_{j}$ and $\textbf{x}_{k}$. One can eliminate the center of mass coordinate and describe the system in terms of two relative Jacobi coordinates. 
The three sets of such coordinates are given by
\begin{equation}
 \mbox{\boldmath$r$}_{i} = \textbf{x}_{j} - \textbf{x}_{k}\quad\text{and}\quad
 \mbox{\boldmath$\rho$}_{i} = \textbf{x}_i - \frac{m_j\textbf{x}_j+m_{k}\textbf{x}_k}{m_j + m_k} \, ,
\end{equation}
where ($i, j, k$) are taken cyclically among ($1,2,3$). The Faddeev decomposition of the three-body wave function allows one to write it as a sum of three components. In the center-of-mass frame, it reads
\begin{equation}
\Psi(\textbf{x}_{1},\textbf{x}_{2},\textbf{x}_{3}) =\sum^{3}_{i=1} 
\psi^{(i)}(\mbox{\boldmath$r$}_i,\mbox{\boldmath$\rho$}_i) . \end{equation}
Each Faddeev component satisfies the free
Schr\"{o}dinger eigenvalue equation with a three-body energy ($E_3=-\kappa_0^2$). For convenience, we introduce the scaled coordinates $
 \mbox{\boldmath$r$}'_{i} = \sqrt{\eta_i}\, \mbox{\boldmath$r$}_{i}\quad$ and $\quad
 \mbox{\boldmath$\rho$}'_{i} = \sqrt{\mu_i}\, \mbox{\boldmath$\rho$}_{i}$, with reduced masses given by
$\eta_{i} = m_{j}m_{k}/(m_{j}+m_{k})$ and $ \mu_{i} 
= {m_{i}(m_{j}+m_{k})}/M$ where $M = m_i + m_j + m_k$. The three sets of primed coordinates are related to each other by the orthogonal transformations
\begin{eqnarray}
\mbox{\boldmath$r$}'_{j}& =& - \mbox{\boldmath$r$}'_{k}\cos\theta_i + \mbox{\boldmath$\rho$}'_{k}
\sin \theta_i,  \\
\mbox{\boldmath$\rho$}'_{j}& =& - \mbox{\boldmath$r$}'_{k}\sin\theta_i - 
\mbox{\boldmath$\rho$}'_{k}\cos \theta_i,
\end{eqnarray}
where $\tan \theta_i = \left[m_i M/(m_j\ m_k)\right]^{1/2}$. 

For three distinguishable bosons interacting resonantly in all pairs and with vanishing total angular momentum, the corresponding eigenvalue equation can be found by using  hyperspherical coordinates $(R,\alpha_i)$ with 
$r'_i = R \sin \alpha_i$ and $\rho'_i = R \cos \alpha_i $. The solution of each Faddeev component of the wave function is written as~\cite{betpeiPRA}
\begin{widetext}
\begin{eqnarray}
\psi^{(i)}(r'_i,\rho'_i) &=&C^{(i)}   \frac{ K_{ s_n}\left(\sqrt{2} \kappa_0 \sqrt{  r'^{2}_{i}+  \rho'^{2}_{i} } 
\right) }
{ \big(  r'^{2}_{i}+ \rho'^{2}_{i} \big)^{D/2-1/2}}\frac{\sqrt{\sin\big[2 \arctan\left( 
r'_i/\rho'_i\right)\big]}}{\big\{\cos\big[ \arctan\left( r'_i/\rho'_i\right)\big]\ \sin\big[ \arctan\left( 
r'_i/\rho'_i\right)\big]\big\}^{D/2-1/2}}
\nonumber \\
&\times&\left[ P_{s_n/2-1/2}^{D/2-1}\Big\{\cos\big[2 \arctan( 
r'_i/\rho'_i)\big]\Big\}-\frac{2}{\pi}\tan\big[\pi(s_n-1)/2 \big]
 Q_{s_n/2-1/2}^{D/2-1}\Big\{\cos\big[2 
\arctan( r'_i/\rho'_i)\big]\Big\}\right]\, ,\ \ \ \ 
\label{wavefunction}
\end{eqnarray}
\end{widetext}
where the coefficients $C^{(i)}$ give the weight between the different Faddeev components for mass-imbalanced systems, $s_n$ is a scaling exponent and $n$ labels distinct pairs of roots $\pm s_n$, $P_{\nu}^{\mu}(x)$ and $Q_{\nu}^{\mu}(x)$ are the associated Legendre functions and $K_{s_n}$ is the modified Bessel function of the second kind. A finite value for the Faddeev component $\psi^{(i)}$ at $\rho_i =0$ must be imposed so that the angular wave function vanishes at $\alpha_i=\pi/2$, since $\rho_i' = R \cos{\alpha_i}$.

The Bethe–Peierls boundary condition in 
$D$-dimensions is imposed on the total wave function as the distance between any two particles vanishes. In the unitary limit, it reads
\begin{equation}
\label{eq:BP3B}
\hspace{-0.2cm}\left[\frac{\partial}{\partial r_i}  
r_{i}^{\frac{D-1}{2}}\Psi(\mbox{\boldmath$r$}_i,\mbox{\boldmath$\rho$}_i)
\right]\Bigg|_{r_i\rightarrow 0} = \frac{3-D}{2} 
\left[\frac{\Psi(\mbox{\boldmath$r$}_i,\mbox{\boldmath$\rho$}_i)}
{r_{i}^{\frac{3-D}{2}}}\right]\Bigg| _{r_i\rightarrow 0}.
\end{equation}
\begin{widetext}
Applying the Bethe–Peierls boundary condition leads to a homogeneous linear system for the Faddeev weights $C^{(i)}$, with a transcendental dependence on the scaling exponent. Taking the three cyclic permutations of $\{i,j,k\}$, one obtains an expression in which the short-distance limit of the interparticle separation is evaluated analytically, rather than kept as a necessary formal limiting procedure as in previous works
\begin{eqnarray}
C^{(i)}\left[\frac{ \csc \left(\pi D /2\right) \cot \left(\pi s_n / 2\right)\Gamma \left[(D+s_n-1)/2\right]}{(2 \sqrt{2})^{-1}\Gamma \left[(D-2)/2\right] \Gamma \left[(s_n-D+3)/2\right]} \right] + (D -2) \left[ \frac{C^{(j)}\  G(\theta_k)}
{\left(\sin\theta_k \cos\theta_k\right)^{(D-1)/2}}+  \frac{C^{(k)}\  G(\theta_j)}
{\left(\sin\theta_j \cos\theta_j\right)^{(D-1)/2}} 
\right] = 0,\ \ \ 
\label{BPfreealpha}
\end{eqnarray}
\end{widetext}
where $G (\alpha_i)$ is the solution of the hyperangular equation~\cite{betpeiPRA} 
\begin{eqnarray} 
G(\alpha_i) &=& \sqrt{\sin2 \alpha_i}\Big[ P_{s_n/2-1/2}^{D/2-1}\,(\cos2\alpha_i)\nonumber \\
&-& \frac{2}{\pi}\tan\big[\pi(s_n -1)/2\big] Q_{s_n/2-1/2}^{D/2-1}\,(\cos2\alpha_i)\Big],\,\hspace{0.8cm}
\label{Eq:AngSol}
\end{eqnarray}
and $\Gamma(z)$ is the gamma function defined for
all complex numbers z. The scaling exponents $s_n$ and the weights of the different Faddeev components are obtained by solving the characteristic transcendental equation of the system.


\subsection{Momentum space}

We review the derivation of the 
three-body wave function of a mass-imbalanced system at unitarity in momentum space. For that, we consider $\textbf{k}_i$, $\textbf{k}_j$ and $\textbf{k}_k$ as the momenta of each particle in the rest frame. The Jacobi momenta of the pairs and from one particle to the center of mass of the other two are given, respectively, by
\begin{eqnarray}
\frac{\textbf{p}_i}{\eta_i} = \frac{\textbf{k}_j}{m_j} - \frac{\textbf{k}_k}{m_k}\ \ \text{and}\ \
\frac{\textbf{q}_i}{\mu_i} = \frac{\textbf{k}_i}{m_i}-\frac{\textbf{k}_j+\textbf{k}_k}{m_j+m_k}, \ \ 
\end{eqnarray}
where $(i,j,k)$ are taken cyclically. The Jacobi momenta $\textbf{p}_i$ and $\textbf{q}_i$ are related to each other as
\begin{eqnarray}
    \textbf{q}_j &=& \textbf{p}_i + \frac{m_j}{m_j+m_k}\textbf{q}_i,\nonumber \\
    \textbf{q}_k &=& \textbf{p}_i - \frac{m_k}{m_k+m_j}\textbf{q}_i.
\end{eqnarray}

In order to write the total bound trimer wave function 
\begin{equation}
\Psi(\textbf{k}_{i},\textbf{k}_{j},\textbf{k}_{k}) = \frac{\chi^{(i)}(\textbf{q}_i)+\chi^{(j)}(\textbf{q}_j)+\chi^{(k)}(\textbf{q}_k)}{E_3 + H_0}, 
\label{eq:fad}
\end{equation}
where $H_0$ is the free Hamiltonian in momentum space, we need to compute the spectator function, $\chi^{(i)}(\textbf{q}_i)$. For that, we need the asymptotic form of 
the Faddeev wave function, Eq.~\eqref{wavefunction}, which is written as
\begin{eqnarray}
\psi^{(i)}\left(\rho'_i,r'_i \right)&\underset{r'_i\to 0}{=}&\mathcal{C}^{(i)} \frac{\sqrt{2} \left[1+\cot \left(D \pi/ 2\right) \cot \left(s_n \pi/2  \right)\right]}{\Gamma \left(2-D/2\right)} \nonumber \\
&\times& \hspace{-0.2cm} \ {r'_i}^{2-D}\frac{K_{s_n}\left(\sqrt{2}\kappa_0 \rho'_{i} \right)}{\rho'_i}, 
\label{wfrlimit}
 \end{eqnarray}
where $\Gamma(z)$ is the gamma function defined for all complex numbers $z$, 
except for the non-positive integers.~This condition restricts the validity of our results to the interval $2\leq D <4$. 

The spectator function $B^{(i)}(\rho_i)$ can be obtained by projecting the full three-body Schrödinger equation onto the Faddeev component $\psi^{(i)}(\rho'_i,r'_i)$, which satisfies a free Schrödinger equation with an inhomogeneous contact source term
\begin{equation}
\left[\nabla_{r'_i}^{2}+\nabla_{\rho'_i}^{2}- 2\kappa_0^2 \right] \psi^{(i)}(r'_i,\rho'_i)= \delta(r'_i)B^{(i)}(\rho'_i)\,.
\label{freeschroe}
\end{equation}
We substitute Eq.~\eqref{wfrlimit} in~\eqref{freeschroe}, which, in the limit $r'_{i}\to0$, is given by
\begin{eqnarray}
B^{(i)}(\rho'_i)&=&\mathcal{C}^{(i)}\frac{2^{3/2} \pi^{D/2}\left[1+\cot \left(D\pi/ 2\right) \cot \left(s_n \pi/2  \right)\right]}{\Gamma(D/2)\Gamma \left(2-D/2\right)} \nonumber \\
&\times&\frac{K_{ s_n}\left(\sqrt{2}\kappa_0 \rho'_{i} \right)}{\rho'_i}.
 \label{specfuncrho}
\end{eqnarray}
Taking the $D$-dimensional FT
\begin{equation}
\int d^{D}\rho'_i \exp(-i \textbf{q}'_i.\bm{\rho}'_i)B^{(i)}(\rho'_i) = \chi^{(i)}(q'_i)\,,
\end{equation}
one can write the spectator function in momentum space ($q'_i=  q_i/\sqrt{\mu_i}$) as
 \begin{eqnarray}
 \chi^{(i)}(q'_i) &=&\mathcal{C}^{(i)}\mathfrak{F}_{(D,s_n)}
\kappa_{0}^{1-D} \nonumber \\
&\times &
H_2  \tilde{F}_1 \left(\mathcal{F}_{(D,s_n)}^{-} ,\mathcal{F}^{+}_{(D,s_n)},\frac{D}{2},-\frac{q_i^{\prime 2}}{2\kappa_0^2}
 \right),\hspace{0.8cm} 
 \label{regularspec}
\end{eqnarray} 
where $\mathcal{F}_{(D, s_n)}$ is given in Eq.~\eqref{DefF} and $H_2 \tilde{F}_1(a,b,c,z)$ is the regularized hypergeometric function with $\mathcal{F}_{(D, s_n)}^{\pm} \equiv (D-1\pm s_n)/2$.


\section{Momentum distribution in $D$-dimensions}
\label{appB}

We illustrate our study focusing on the momentum distribution of the particle $B$. By means of the Faddeev
decomposition shown in Eq.~\eqref{eq:fad}, the momentum density in Eq.~\eqref{totdenB} can be split into nine terms, which, because of the symmetry between the two identical particles $A$, are reduced to
 \begin{equation}
 \label{4sum}
 n_B(q_B) = n_1(q_B) + n_2(q_B) + n_3(q_B) + n_4(q_B),
 \end{equation}
where
\begin{eqnarray}
n_{1}(q_B)= \lvert \chi^{(B)}(q_B) \rvert^{2} \int d^{D}p_B \frac{1}{\left(E_3 +  p_{B}^{2}+q_{B}^{2}/2\mu_B\right)^{2}},   \nonumber \\
 \label{n1}
 \end{eqnarray}
 \begin{eqnarray}
 n_{2}(q_B) = 2 \int d^{D}p_B \frac{\lvert \chi^{(A)}(\lvert \textbf{p}_B -\textbf{q}_B/2 \rvert) \rvert^{2}}
 { \left( E_3 + p_{B}^{2}+ q_{B}^{2}/2\mu_B \right)^{2} }, \hspace{1.4cm}  
 \label{n2}
\end{eqnarray}
 \begin{eqnarray}
 n_{3}(q_B) &=& 2  \chi^{(B)}\overset{*}{(}q_B) \int d^{D}p_B \frac{\chi^{(A)}(\lvert \textbf{p}_B - \textbf{q}_B/2  \rvert )  }
 { \left( E_3 + p_{B}^{2} + q_{B}^{2}  /2\mu_B \right)^{2} },\nonumber \\
 &+& {\rm c.c.},
 \label{n3}
 \end{eqnarray}
 \begin{eqnarray}
 n_{4}(q_B) &=& \int d^{D}p_B \frac{\chi^{(A)}\overset{*}{(}\lvert \textbf{p}_B-\textbf{q}_B/2 \rvert)
 \chi^{(A)}(\lvert \textbf{p}_B + \textbf{q}_B/2  \rvert )  }
 {\left( E_3 + p_{B}^{2} + q_{B}^{2}  /2\mu_B\right)^{2}} \nonumber \\
 &+& {\rm c.c.}.
 \label{n4}
 \end{eqnarray}

Now, we turn our attention to the leading and subleading contributions of the high-momentum tail of the single-particle momentum distribution from which the two-body contact and three-body tail coefficients are extracted. 

We start by examining the contribution $n_1(q_B)$, Eq.\eqref{n1}, which is straightforward to calculate
 \begin{eqnarray}
 n_{1}(q_B) &=&  \frac{\lvert \chi^{(B)}(q_B) \rvert^{2}}{q_{B}^{4-D}} \mathcal{S}_{D}    \frac{\pi}{4}\csc\left( \frac{D\pi}{2} \right) (2-D)\nonumber \\ &\times&\left(2\mu_B\right)^{2-D/2} \, ,
 \label{eq:n1desenvolved}
 \end{eqnarray} 
where $\mathcal{S}_{D}$ is the area of a $D$-dimensional sphere. 

The second contribution, $n_2(q_B)$, can be computed 
from Eq.~\eqref{n2} by making the change of variables $\textbf{p}_B-\textbf{q}_B/2=\textbf{q}_A$, that results in 
\begin{equation}
 n_{2}(q_B) = 2 \int d^{D}q_A \frac{\lvert \chi^{(A)}(q_A) \rvert^{2}}
 { \left(  q_A^{2}+ \textbf{q}_A.\textbf{q}_B + q_{B}^{2}  \mu_A/\mu_B \right)^{2} }\, .
 \end{equation}
In order to identify the leading order term, we perform the 
manipulation~\cite{castindensity,braatendensity}

\begin{eqnarray}
    &&\left[q_A^{2}+ \textbf{q}_A\cdot\textbf{q}_B + q_{B}^{2}  \mu_A/\mu_B \right]^{-2} = \left(\mu_B/\mu_A \right)^2q_B^{-4}\nonumber \\
    &&+\left[(q_A^{2}+ \textbf{q}_A\cdot\textbf{q}_B + q_{B}^{2}  \mu_A/\mu_B)^{-2} - \left(\mu_B/\mu_A \right)^2q_B^{-4}\right],\hspace{1.0cm}
\end{eqnarray}
we can then write
\begin{eqnarray}\label{eq:n2}
 n_{2}(q_B) &=& 2 \int d^{D}q_A\lvert \chi^{(A)}(q_A) \rvert^{2}\nonumber \\
 &\times&\left[ \frac{1}
 { \left( q_A^{2}+ \textbf{q}_A.\textbf{q}_B + q_{B}^{2}  \mu_A/\mu_B \right)^{2} }-\left(\frac{\mu_B}{\mu_A }\right)^2\frac{1}{q_B^4}\right] \nonumber \\
 & +& \frac{C_2}{q_B^{4}}\, ,
 \end{eqnarray}
where $C_2$ is the two-body contact and has dimension  
$(\text{length})^{D-4}$ and therefore  scales as $C_2\propto \kappa_0^{4-D}$. The two-body contact parameter is given by
 \begin{eqnarray}
 C_{2}
= 2\left(\frac{\mu_B}{\mu_A }\right)^2\mathcal{S}_D \int^\infty_0 dq_A \, q_A^{D-1}\lvert \chi^{(A)}(q_A) \rvert^{2}\,.  \label{c2}
 \end{eqnarray}
In spherical coordinates, Eq.~\eqref{eq:n2} reads
\begin{eqnarray}
 &&n_{2}(q_B) = \frac{2(2\pi)}{q_{B}^{4}}\prod_{k=1}^{D-3}\int_0^{\pi} d\theta_k\sin^{k}\theta_k \nonumber \\
 &&\times\int^\infty_0 dq_A\   q_A^{D-1}  | \chi^{(A)}(q_A) |^{2}
\int_0^{\pi} d\theta\ \sin^{D-2}\theta \nonumber \\
&&\times\left\{\frac{1}{\left[ (q_A/q_B)^{2} + (q_A/q_B)\cos\theta+ \mu_B/\mu_A \right]^{2}}\right. \nonumber \\
&&\left.- \left(\frac{\mu_B}{\mu_A}\right)^2\right\}+\frac{C_2}{q_B^4}\,.
 \end{eqnarray}
\noindent Making the change of variables $q'_A=q_A/ q_B$, one finds
\begin{eqnarray}
n_{2}(q_B) &= & \frac{2\mathcal{S}_D }{q_B^{4-D}}\int^\infty_0 dq'_A \, q_A^{\prime D-1} | \chi^{(A)}(q_B\, q'_A) |^{2}\nonumber \\ &\times& \left(\mathcal{H}(q'_A)-\frac{4 \mathcal{A}^2}{(\mathcal{A}+1)^2}\right)+\frac{C_2}{q_B^{4}},
 \label{eq:n2desenvolved}
 \end{eqnarray}
 with
  \begin{eqnarray}
 \mathcal{H}(y)&\equiv&\frac{4 \mathcal{A}^2 (D-2)}{\mathcal{A}^2 \left(4 y^4+1\right)+\mathcal{A} \left(4 y^2+2\right)+1} \nonumber \\
 &+&\frac{4 \mathcal{A}^2 (3-D) \left(2  y^2\ \mathcal{A}+\mathcal{A}+1\right) \, }{[2 \mathcal{A} (y-1) y+\mathcal{A}+1]^2 [2  y (y+1)\mathcal{A}+\mathcal{A}+1]}\nonumber \\
 &\times& H_2F_1\left(1,\frac{D-1}{2},D-1,\frac{-4 \mathcal{A} y}{2 (y^2-y)  \mathcal{A}+\mathcal{A}+1}\right),\nonumber \\
 \label{eq:H2F1}
 \end{eqnarray}
where $H_2 F_1(a,b,c,z)$ is the hypergeometric function and $\mathcal{A}\equiv m_B/m_A$. 

In the large momentum limit ($q_B\gg \sqrt{2\mu_B E_3}$), after the change of variables $\textbf{p}_B - \textbf{q}_B/2 = \textbf{q}_A$ and the observation that the spectator functions are real, the third contribution, $n_3(q_B$), given in Eq.~\eqref{n3}, can be written as

  \begin{equation}
n_{3}(q_B)=  4\int d^{D}q_A \frac{\chi^{(B)}\overset{*}{(}q_B)\ \chi^{(A)}(q_A )  }
 { \left( q_{A}^{2} +\textbf{q}_A.\textbf{q}_B +q_B^{2}\mu_A/\mu_B \right)^{2} }\,.
 \end{equation}
Changing variables as $q_A/q_B =q'_A$ and integrating in spherical coordinates, we obtain
 \begin{equation}
 n_{3}(q_B)= \chi^{(B)}\overset{*}{(}q_B)\frac{4\mathcal{S}_D}{q_B^{4-D}} \int^\infty_0 \hspace{-.3cm}  dq'_A  \ q_A^{\prime D-1}  \chi^{(A)}(q_B q'_A)\mathcal{H}(q'_A)\,,
 \label{eq:n3desenvolved}
 \end{equation}
where $\mathcal{H}(y)$ is  given by Eq.~\eqref{eq:H2F1}. 

For the contribution $n_4(q_B)$, we observe that there is an argument in the spectator function that can be written as
\begin{eqnarray}
 \lvert \textbf{p}_B\pm\frac{\textbf{q}_B}{2}\rvert = q_B \sqrt{\frac{p_B^{2}}{q_B^{2}}+\frac{1}{4}\pm \frac{p_B}{q_B}\cos\theta},
\end{eqnarray}
where, making the same steps as in the previous contribution, but now changing the variables as $p_B /q_B =  p'_B$, we find
 \begin{eqnarray}
 n_{4}(q_B) &= & \frac{1}{q_B^{4-D}} 4\pi\prod_{k=1}^{D-3}\int_0^{\pi}d\theta_k  \sin^{k}(\theta_k)\nonumber   \\
 &\times&\int^\infty_0 dp'_B  \frac{p_B^{\prime D-1}  }
 { \left( p_{B}^{\prime 2} + 1/2\mu_B \right)^{2} } \int_0^{\pi}d\theta \sin^{D-2}\theta \nonumber  \\
 &\times& \chi^{(A)}\overset{*}{(} q_B \, p'_{B -})\chi^{(A)}(q_B \, p'_{B +})\, ,
 \label{eq:n4desenvolved}
 \end{eqnarray}
 where $ p'_{B\pm}=\sqrt{p^{\prime 2}_B+\frac14\pm p'_B\cos\theta}$.

\begin{widetext}
\section{Two-body contact and three-body tail coefficients for different mass configurations}
\label{appC}

To complement the analysis presented in Fig.~6 of Ref.~\cite{unatomic2}, we include in Fig.~\ref{fig10} the values of the three-body tail coefficients evaluated at the critical dimension $D_c$ for representative mass configurations. These points are superimposed on the existing curves using distinct blue markers: empty squares for $C_3''$, empty circles for $C_3'$, and solid circles for $C_3$. The corresponding systems are explicitly indicated in the figure.

\begin{figure}[h]
\includegraphics[width=8.5cm]{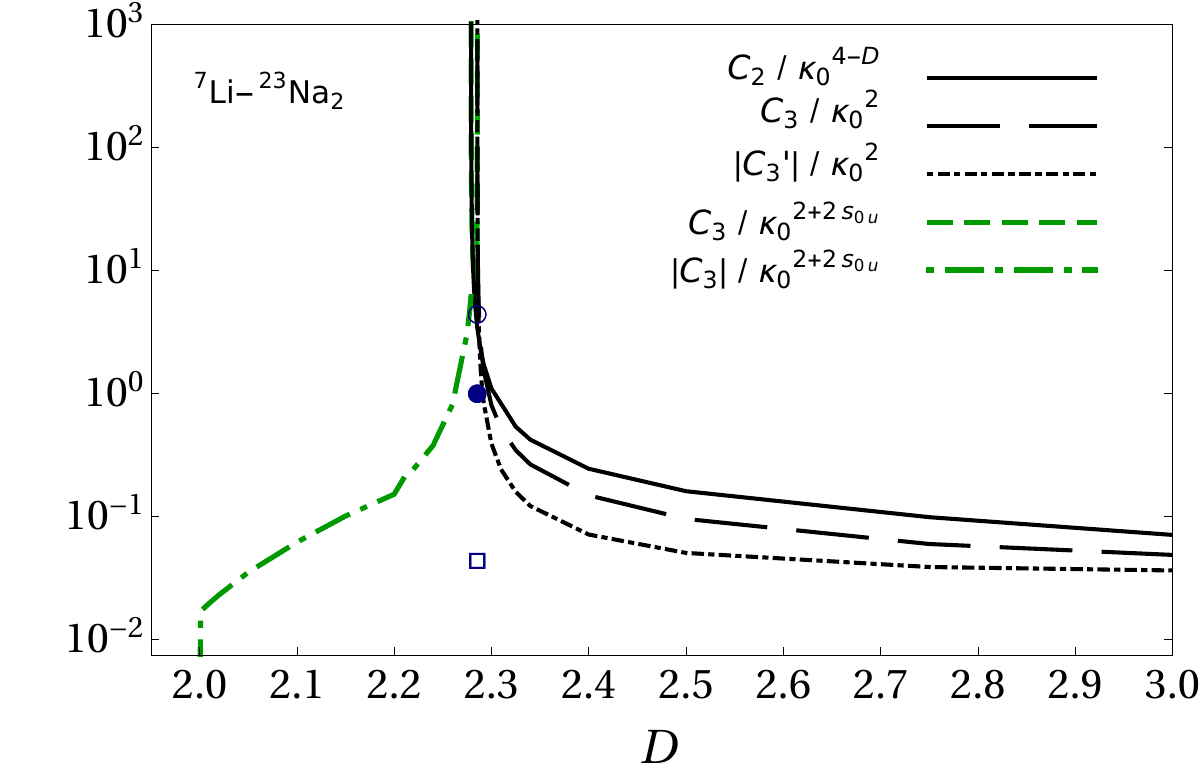}
\includegraphics[width=8.5cm]{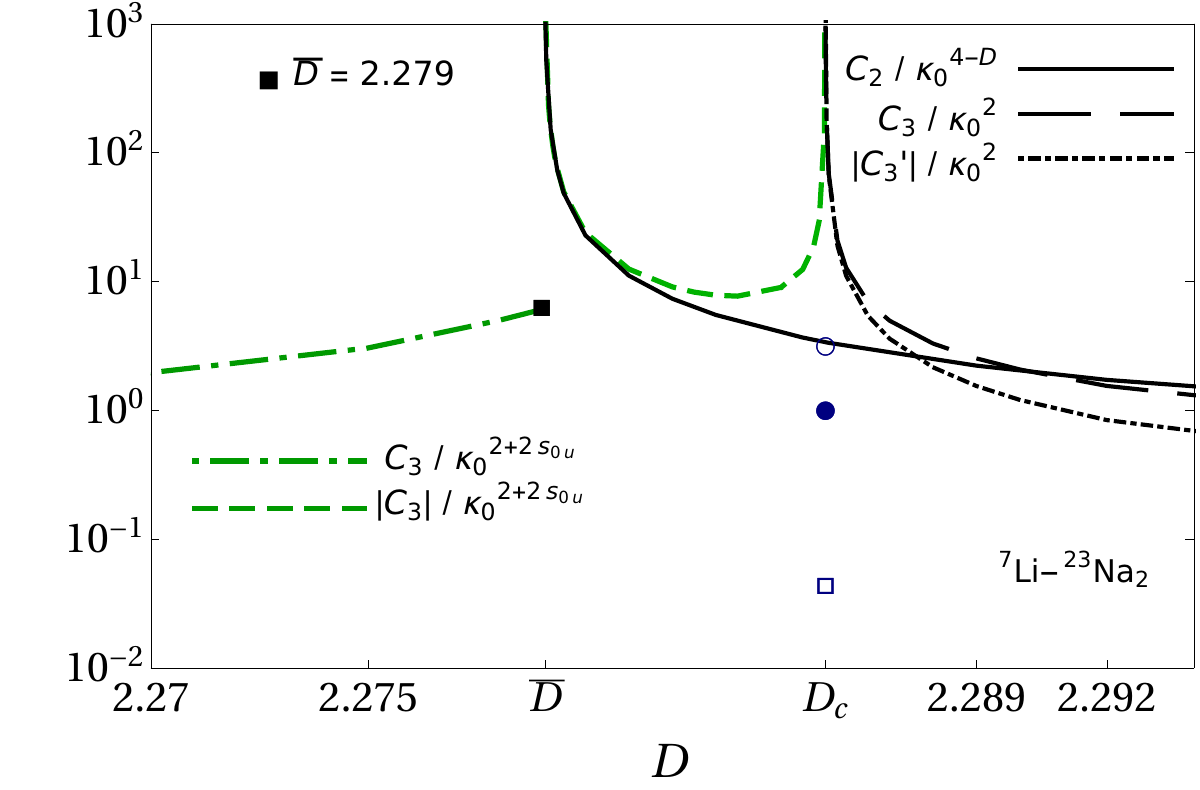}
\includegraphics[width=8.5cm]{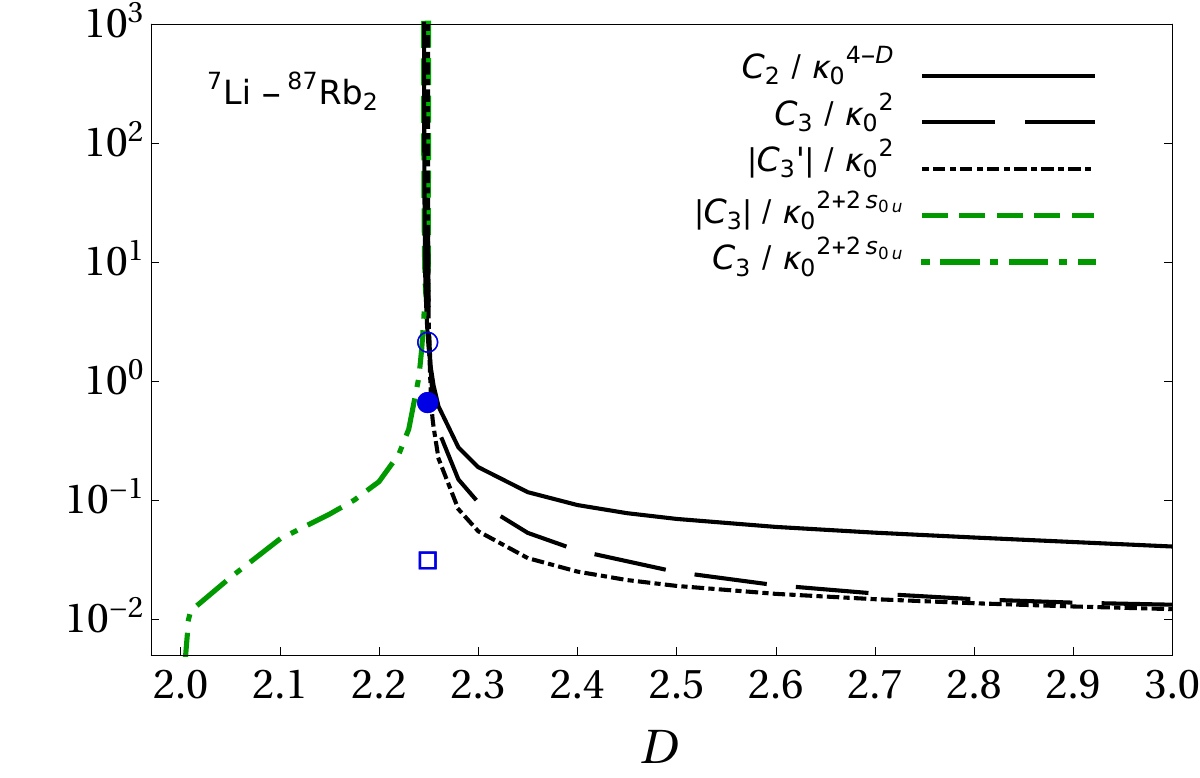}
\includegraphics[width=8.5cm]{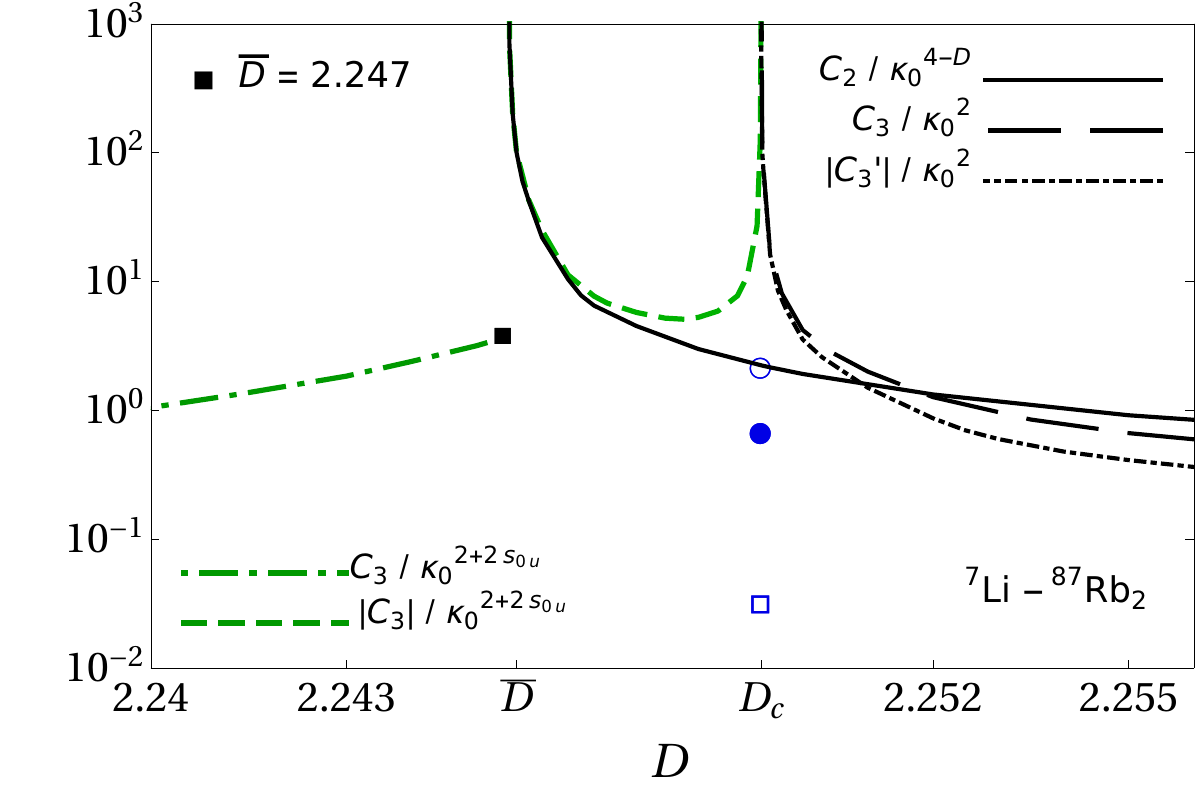}
\includegraphics[width=8.5cm]{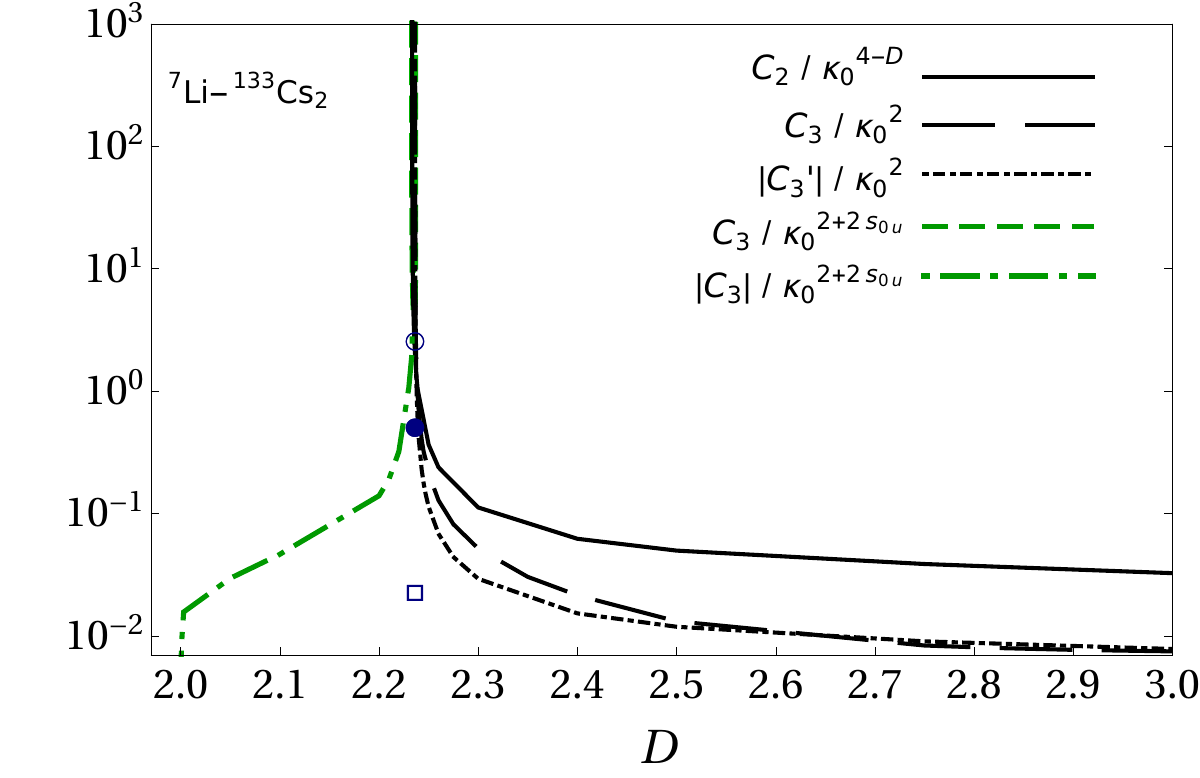}
\includegraphics[width=8.5cm]{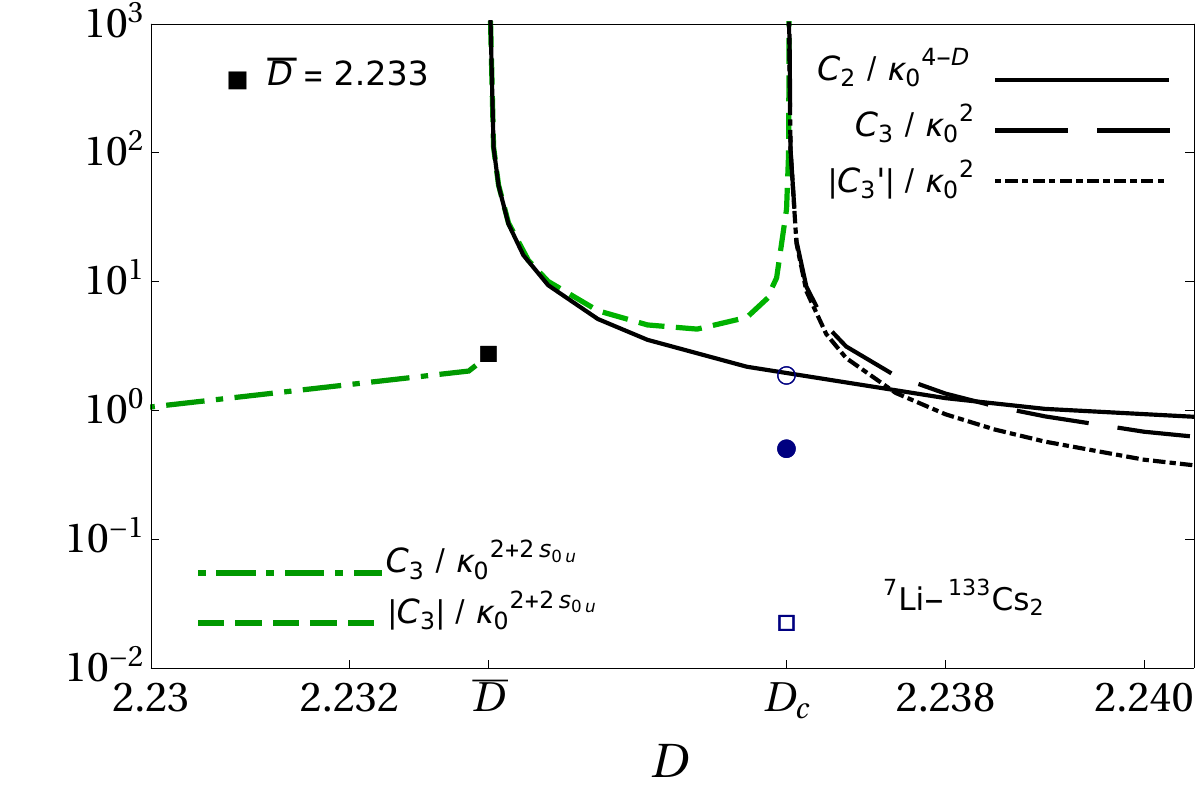}
\caption{The two-body contact and three-body tail coefficients as functions of the noninteger dimension $D$. Left panels: from $D=2$ to $D=3$. Right panels: zoom in around the ISR region.} 
\label{fig10}
\end{figure}
    
\end{widetext}


\end{document}